\documentclass[12pt]{article}
\usepackage{amsmath}
\usepackage{amssymb}
\usepackage{graphicx}% Include figure files
\usepackage{dcolumn}% Align table columns on decimal point
\usepackage{bm}% bold math

\newcommand{\prt}{\partial}

\newcommand{\la}{\lambda}
\newcommand{\K}{\mathrm{K}}
\newcommand{\E}{\mathrm{E}}

\begin{document}

\title{\bf
Refraction of dispersive shock waves}

\author{
G.A. El $^{1,*}$,  V.V. Khodorovskii $^2$ and A.M. Leszczyszyn $^1$   \\ \\
$^1$ Department of Mathematical Sciences, Loughborough University,\\
Loughborough,  LE11 3TU, UK \\ \\
$^2$ St. Petersburg Institute of Judaic Studies,  \\ 2, Ruzovskaya str., St. Petersburg, 190013, Russia}
\date{}
\maketitle
\begin{abstract}
We study a dispersive counterpart of the classical gas dynamics problem of the interaction  of a shock wave with a counter-propagating simple rarefaction wave often referred to as the shock wave refraction. The refraction of a one-dimensional dispersive shock wave (DSW) due to its head-on collision with the centred rarefaction wave (RW) is considered in the framework of defocusing nonlinear Schr\"odinger  (NLS) equation.  For the integrable cubic nonlinearity case we present a full asymptotic description of the DSW refraction by constructing appropriate exact solutions of the Whitham modulation equations in Riemann invariants. For the NLS equation with saturable nonlinearity, whose modulation system does not possess Riemann invariants, we take advantage of the recently developed method for the DSW description in non-integrable dispersive systems to obtain main physical parameters of the DSW refraction. The key features of the DSW-RW interaction  predicted by our  modulation theory analysis are confirmed by direct numerical solutions of the full dispersive problem.

\end{abstract}
\footnotetext[1]{Corresponding author. email: g.el@lboro.ac.uk; tel/fax: +44 15092222869/+44 1509225969}

\section{Introduction}

Recent developments  of experimental techniques  of cold-atom and laser physics and observations of a number of superfluid and optical counterparts of classical hydrodynamic phenomena such as solitons, shock waves, rarefaction waves, vortex streets etc. (see e.g. \cite{cornell05, ha06, fleischer07,  conti07, ea07,  karman10, amo11}) stimulated the growing interest in the mathematical methods and results of dispersive hydrodynamics --- the theory of  multiscale nonlinear flows in media with dispersive (rather than dissipative) mechanisms of regularization of breaking singularities.
Central to dispersive hydrodynamics is the theory of dispersive shock waves (DSWs),  which  represent expanding nonlinear wavetrains connecting two different hydrodynamic states and replacing, in conservative continuous media, the classical viscous shocks (see \cite{scholarpedia} and references therein).  Owing to their rich dynamics and fundamental physical nature, the DSWs have recently become an object of very active theoretical and experimental investigations, most notably in Bose-Einstein condensates (BECs) (see, e.g., \cite{cornell05, ha06, chang08}), where these waves represent a striking manifestation of quantum statistics on a macroscopic scale. Another area of active modern DSW research is nonlinear optics (see \cite{kod99, fleischer07, conti07, assanto}).

While the dynamics of isolated DSWs  have been studied in numerous works since the pioneering paper \cite{gp74} by Gurevich and Pitaevskii, their interaction behaviour has  begun to be investigated relatively recently. One can distinguish two prototypical problems arising in this connection: interaction two DSWs and interaction of a DSW with a simple (rarefaction) wave.
Both problems admit full analytical description in the framework of the  Whitham modulation theory associated with integrable wave equations, however, the former problem  involves complicated analysis of nonlinear multiphase wavetrains (see e.g. \cite{grava} for the KdV equation) so in applied problems one is usually better off finding the corresponding modulation solutions numerically (see \cite{abh09}, \cite{bk06}, \cite{ha07}). The latter problem of the DSW-RW interaction, on the contrary, involves only single-phase dynamics, so one could hope for a relatively simple  effective analytic description of the interaction.

In  \cite{abh09} a complete classification of {\it unidirectional}  interactions of DSWs and RWs in weakly dispersive flows was made using the analytical inverse scattering transform (IST) solutions for the KdV equation and numerical solutions of the KdV-Whitham equations. This classification has revealed  certain similarities as well as fundamental differences between classical and dispersive-hydrodynamic overtaking shock wave-rarefaction wave interactions (see also \cite{eg02} for the analytical modulation solutions describing  some of the KdV overtaking DSW-RW interactions). In many physical settings, however, one has to deal with {\it bi-directional} (head-on) wave collisions which cannot be captured by the KdV type models and should be studied in the framework of  appropriate two-wave equations. Such a bi-directional  DSW-RW interaction represents a dispersive counterpart of the classical gas dynamics problem which is often referred to as the ``shock wave refraction''.

When a one-dimensional viscous shock wave (SW) undergoes a head-on collision  with a  rarefaction wave, the parameters of two waves alter so that the long-time output of such an interaction consists of a new pair of SW and RW propagating in opposite directions. Since the SW speed changes from one constant value to another as a result of its propagation through the finite  RW region with varying density and velocity, the interaction diagram in the $(xt)$-plane could be naturally interpreted as the SW refraction on the RW. As a matter of fact, the SW refraction can be observed in two-dimensional stationary flows  where the effect acquires its direct geometrical significance as deflection of the SW from its original propagation direction accompanied by the change of its strength and speed.

Refraction of SW's has been the subject of many  gas and fluid dynamics investigations (see, e.g.
the original wartime report \cite{cf43} by Courant and Friedrichs and some of the well-known research papers \cite{moses, AYa, hasimoto, Roz} as well as classical monographs \cite{cf48}, \cite{RY}, \cite{ovs}). It  must be said, however, that, while the qualitative features of the SW refraction process have been understood very well, its analytical description is seriously hindered due to the presence of the varying entropy region between the refracted SW and RW. As a result, the system of equations describing the head-on SW-RW interaction turns out to be so complicated that numerical solution becomes in most cases the only available resort.

In dispersive compressible dissipationless flows the entropy does not change and, in contrast to viscous gas dynamics, the bidirectional DSW-RW interaction can be described analytically in terms of  solutions of the  Whitham modulation  equations \cite{whitham74} associated with the original dispersive-hydrodynamic system and governing  slow variations of the wave parameters (amplitude, wavenumber, mean etc.) on the scale much larger than the medium typical coherence length.

In this paper, we perform an analytical study of the head-on DSW-RW interaction in the framework of the nonlinear Schr\"odinger equation with defocussing, which is a
standard mathematical model in nonlinear optics and condensed matter physics (see e.g. \cite{ka03}, \cite{ps03}).  Thus, apart from the obvious theoretical significance as a dispersive counterpart of a classical gas dynamics problem, the theory of the DSW refraction in nonlinear Schr\"odinger flows is fundamental to the understanding of `dispersive-hydrodynamic' flow interactions in superfluids and nonlinear optical media.
For the case of cubic nonlinearity the NLS equation is a completely integrable system  and a full asymptotic description of the DSW-RW
interaction becomes possible owing to the availability of  exact solutions of the NLS-Whitham equations describing slow variations of the rapidly oscillating wave field in the interaction zone.
The key element of the analytical construction  is the  mapping of the two-component
reduction of the NLS-Whitham system to the classical linear Euler-Poisson-Darboux (EPD) equation. This mapping was introduced for the KdV-Whitham system in \cite{kudshar91} \cite{gke91}, \cite{gke92} and \cite{tian93}; and for the NLS equation in \cite{gke92}.
Remarkably,
the same EPD equation  describes, on the hodograph plane, the interaction of two nonlinear simple waves in ideal shallow-water dynamics -- see, e.g. \cite{whitham74}.

Along with the study of the DSW refraction in Kerr media described by the integrable NLS equation,
we also undertake a similar investigation  of the DSW-RW interaction in the framework of  the NLS equation with saturable nonlinearity (sNLS), which represents a standard model for the  optical beam propagation in photorefractive crystals (see, e.g.   \cite{gatz91}, \cite{christ95}, \cite{ka03}). The photorefractive systems have been recently used for the modelling dispersive-hydrodynamic flows in BECs by means of an all-optical setting  \cite{fleischer07} so the quantification of the contribution of the saturation effects to the `superfluid' dynamics of light is important for the comparison with BEC experiments.

 The sNLS equation is not integrable by the inverse scattering transform, so the associated Whitham system does not possess Riemann invariants. As a result, this system cannot be reduced to the EPD equation and the analytic method employed for the description of the DSW refraction in the cubic nonlinearity case is not applicable to the sNLS equation. To tackle the sNLS refraction problem analytically, we take advantage of the  approach to the dispersive Riemann problem treatment in non-integrable conservative systems developed in \cite{el05}. This has enabled us to derive analytically the key parameters  of the refracted DSW,  as well as the DSW refraction, acceleration and amplification coefficients as functions of the initial data and the saturation parameter $\gamma$. We note that the theory of propagation of simple photorefractive DSWs was developed in paper \cite{egkk07}, which contains  some detailed explanations of the application of the method of \cite{el05} to the sNLS equation. In the present paper, we extend the results of \cite{egkk07} to describe the photorefractive DSW-RW interaction.  In particular, we show that for a broad range of parameters the photorefractive DSW-RW interaction is asymptotically ``clean'', i.e. is not accompanied by the generation of new DSWs or RWs.

 The direct numerical simulations for NLS and sNLS equations (using standard split-step Fourier method -- see, e.g., \cite{Numerical}) confirm  all the key features of the bidirectional DSW-RW interaction, predicted by our modulation analysis.  We stress, however, that, while we perform some basic comparisons for the typical behaviours of the key physical parameters,   a systematic validation of the obtained solutions in the framework of full dispersive problem is beyond the scope of the present paper and would require, in particular, a detailed comparison with the numerical solutions of the small-dispersion limit of the NLS equation (see, e.g., \cite{gk07} for the corresponding analysis for the KdV equation).  At the same time we would like to mention that there have been a number of comparisons between the solutions of the NLS-Whitham systems and direct numerics for the NLS equation in recent literature (see e.g. \cite{kgk04}, \cite{ha06}, \cite{egkk07}, \cite{wing09}), all of them showing a very good agreement, so we have some confidence in the relevance of our modulation solutions to the properties of the full dispersive DSW-RW interaction dynamics.

\section{DSW refraction in Kerr media: formulation of the problem}
We first formulate the problem for the defocusing NLS equation with cubic (Kerr) nonlinearity
\begin{equation}\label{3-1}
    i\epsilon \psi_t+\frac{\epsilon^2}{2}\psi_{xx}-|\psi|^2\psi=0 ,
\end{equation}
where $\psi$ is a complex valued function and $\epsilon$ is a dimensionless dispersion parameter (coherence length). Using the Madelung transformation $\psi \mapsto (n,u)$
\begin{equation}\label{3-2}
    \psi(x,t)=\sqrt{n(x,t)}\exp\left(\frac{i}{\epsilon}\int^x u(x',t)dx'\right)\, ,
\end{equation}
where $n(x,t)>0$ and $u(x,t)$ are real-valued functions,
we represent the NLS equation (\ref{3-1})
in the ``dispersive-hydrodynamic'' form
\begin{equation}\label{NLS}
\begin{split}
    n_t+(nu)_x=0,\\
 u_t+uu_x+n_x+\epsilon^2\left(\frac{n_x^2}{8n^2}
   -\frac{n_{xx}}{4n}\right)_x=0,
   \end{split}
\end{equation}
with the `fluid' density $n$ and velocity $u$.

The dispersionless (classical) limit of system (\ref{NLS}) is obtained by setting  $\epsilon=0$
and is nothing but the system of ideal shallow-water equations
\begin{equation}\label{eq21}
 n_t+(nu)_x=0,\quad u_t+uu_x+n_x=0\, ,
\end{equation}
which can be represented in the diagonal form
\begin{equation}\label{er}
\frac{\partial \la_{\pm}}{\partial t} + V_{\pm}(\la_+, \la_-)
\frac{\partial \la_{\pm}}{\partial x}=0\, ,
\end{equation}
with the Riemann invariants
\begin{equation}\label{eq20}
 \la_\pm=\frac12{u}\pm\sqrt{n}\,
\end{equation}
and the characteristic velocities
\begin{equation}\label{V}
V_+=\frac{3}{2} \la_++ \frac12{\la_-} \, , \qquad V_-=\frac32
{\la_-}+ \frac12{\la_+} \, .
\end{equation}

To study the bidirectional (head-on) interaction of a DSW  and a rarefaction wave (RW) we consider the following  configuration. Let at some moment of time say $t=t_c$, a simple right-propagating DSW confined to the expanding region $x_1^-(t)< x < x_1^+(t)$ and a simple left-propagating RW located at $x_2^-(t)< x < x_2^+(t)$, be separated by an undisturbed flow region $x_1^+(t)<x<x_2^-(t)$ with $n=1$ and $u=0$ (see Fig. 1).
\begin{figure}[h]
\begin{center}
\includegraphics[width=8cm]{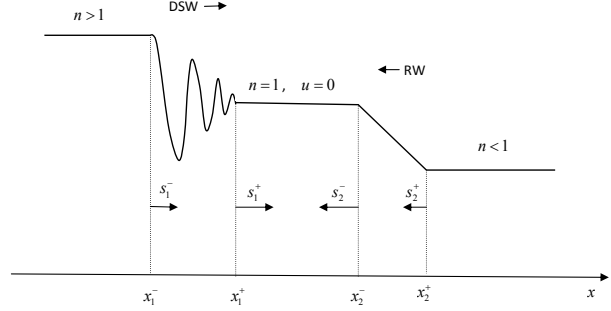}
\caption{Sketch of the density profile in the NLS flow prior to head-on DSW-RW interaction}
\label{fig1}
\end{center}
\end{figure}
Without much loss of generality one can  assume that the DSW and RW are both centred in the $(x,t)$-plane at $(0,0)$ and $(0,l)$ respectively, so that $x_1^{\pm}=s_1^\pm t$, $x_2^{\pm}=l+s_2^\pm t$, where
 $s_1^+>s_1^->0$, $s_2^-<s_2^+<0$ are the speeds of the respective DSW and RW edges. We also assume that $l \gg 1$.

The following transition conditions must be satisfied across the DSW and RW respectively (see \cite{gk87}, \cite{eggk95}):
\begin{eqnarray}\label{}
\la_-(x^-_1, t_c) &=& \la_-(x^+_1, t_c) =-1\qquad \hbox{simple right-propagating DSW transition} \label{trDSW}\\
\la_+(x^-_2, t_c) &=& \la_-(x^+_2, t_c) =1 \qquad \hbox{simple left-propagating RW transition} \label{trRW}
\end{eqnarray}
The transition conditions (\ref{trDSW}), (\ref{trRW}) imply that
the described above flow configuration can be realised as a result of the evolution of the  initial flow profile $n(x,0)$, $u(x,0)$ specified in terms of the shallow-water Riemann invariants $\la_{\pm}$ (\ref{eq20}) having the jumps of different polarity shifted with respect to one another by a distance $l$ (see Fig 2a):
\begin{figure}[h]
\begin{center}
\includegraphics[width=6cm]{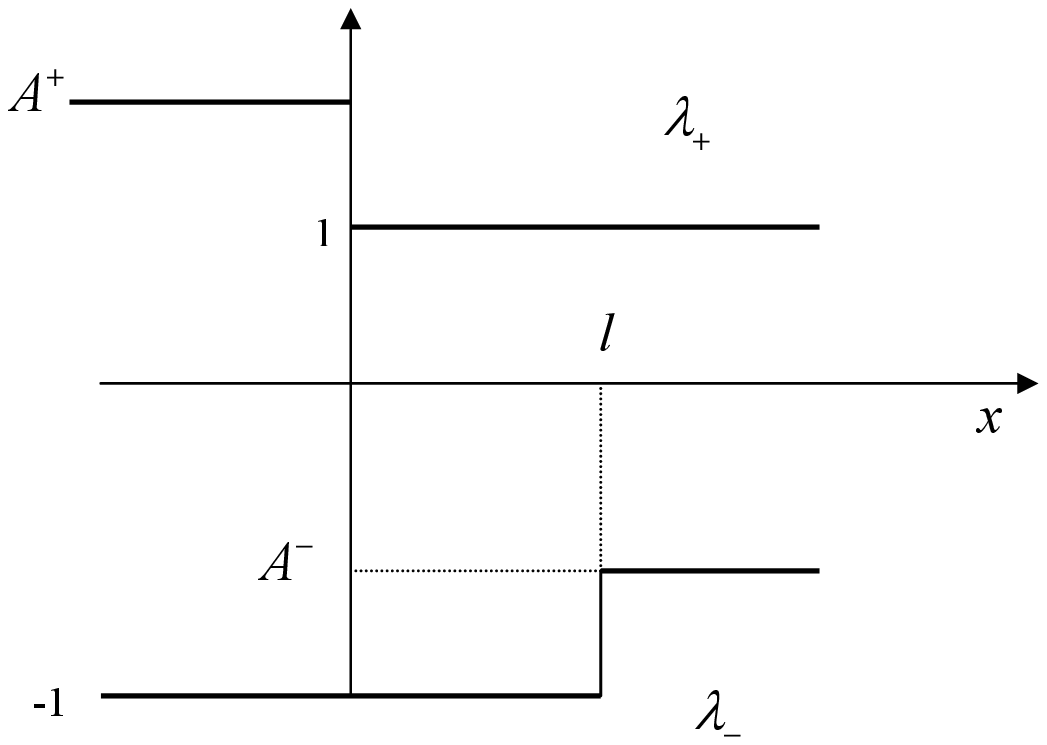}\qquad \qquad  \quad \includegraphics[width=5cm]{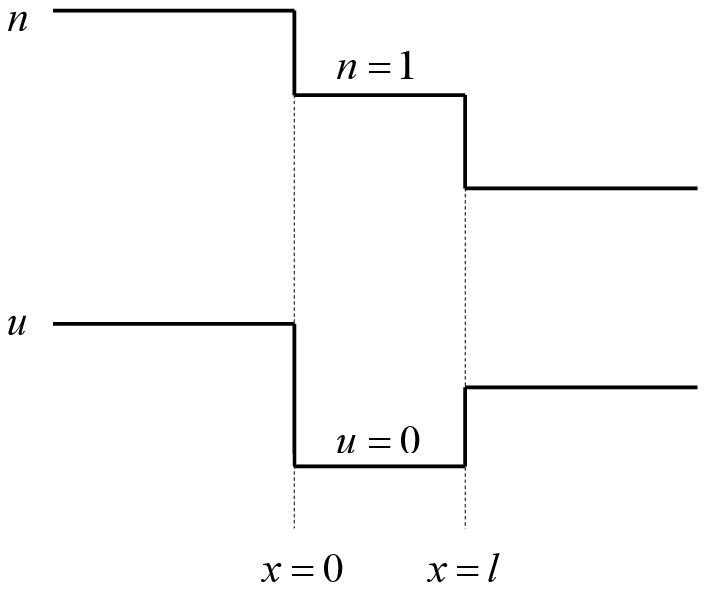}
\caption{Initial conditions  for the NLS equation (\ref{NLS}) leading to the head-on DSW-RW interaction. Left: hydrodynamic Riemann invariants $\lambda_{\pm}$ (\ref{ic1}); \ Right: corresponding density $n$ and velocity $u$ distributions (\ref{ic2}).}
\label{fig2}
\end{center}
\end{figure}
\begin{equation}\label{ic1}
\la_+(x,0) =\left\{
\begin{array}{ll}
A^+ &\quad \hbox{for} \quad x <0,\\
 1 & \quad \hbox{for} \quad x>0;
\end{array}
\right.
\qquad  \la_-(x,0)=\left\{
\begin{array}{ll}
 -1 & \quad \hbox{for} \quad x<l, \\
 A^-  &\quad \hbox{for} \quad x >l,
\end{array}
\right.
\end{equation}
where $A^+>1$ and $-1<A^-<1$.

The initial conditions for $n$ and $u$ corresponding  to (\ref{ic1}) are then readily found using (\ref{eq20}) in the form of  piecewise constant distributions (see Fig.~2b)
\begin{equation}\label{ic2}
n(x,0) =\left\{
\begin{array}{ll}
\frac{1}{4}(1+A^+)^2>1 &\quad \hbox{for} \quad x <0,\\
 1 & \quad \hbox{for} \quad 0<x<l, \\
 \frac{1}{4}(1-A^-)^2<1 &\quad \hbox{for} \quad x > l\, ;
\end{array}
\right.
\qquad  u(x,0)=\left\{
\begin{array}{ll}
A^+-1>0 &\quad \hbox{for} \quad x <0,\\
 0 & \quad \hbox{for} \quad 0<x<l, \\
1+A^->0 &\quad \hbox{for} \quad x > l\, .
\end{array}
\right.
\end{equation}
\begin{figure}[htp]
\begin{center}
\includegraphics[width=17cm]{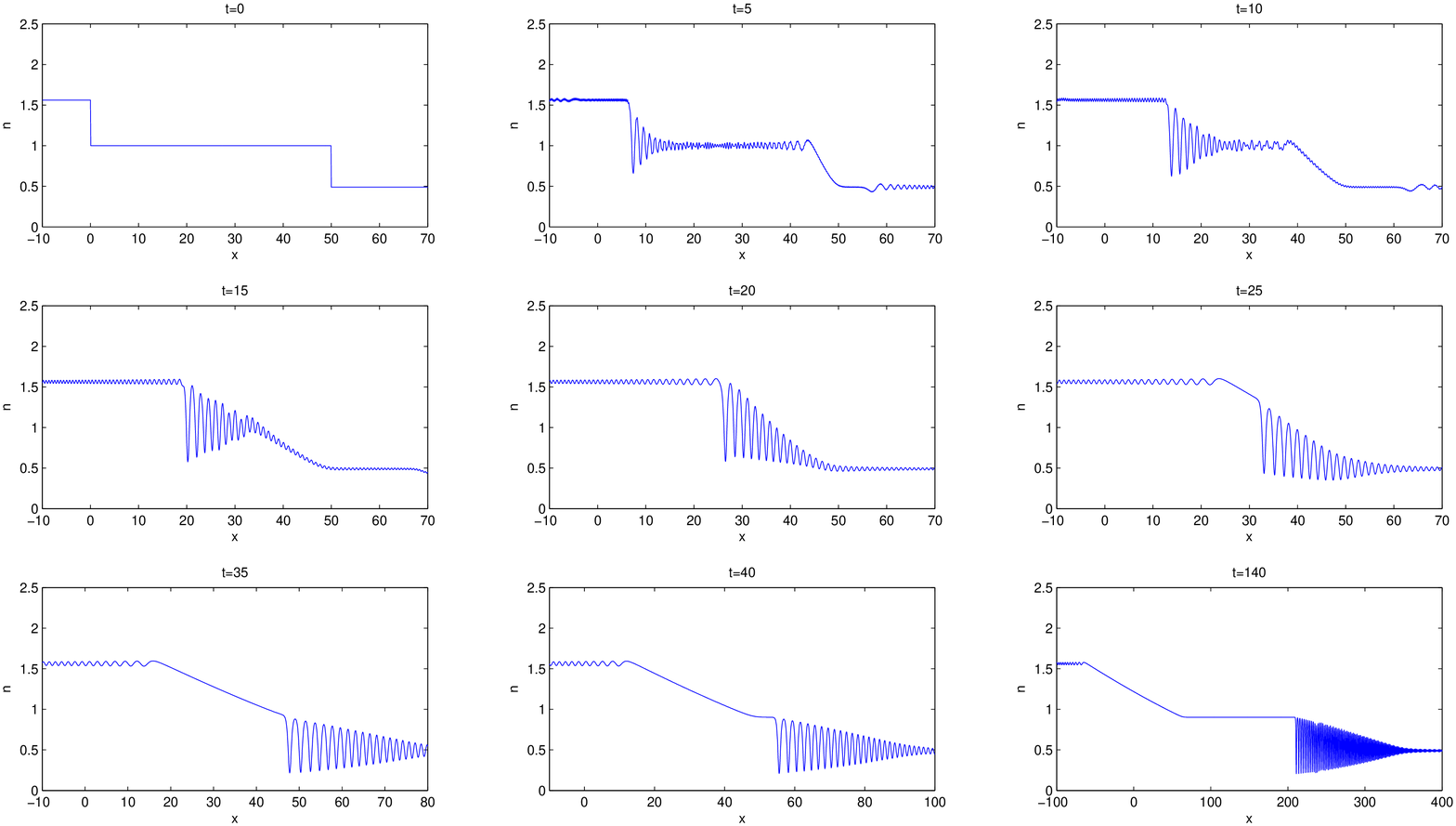}
\vfill
\includegraphics[width=17cm]{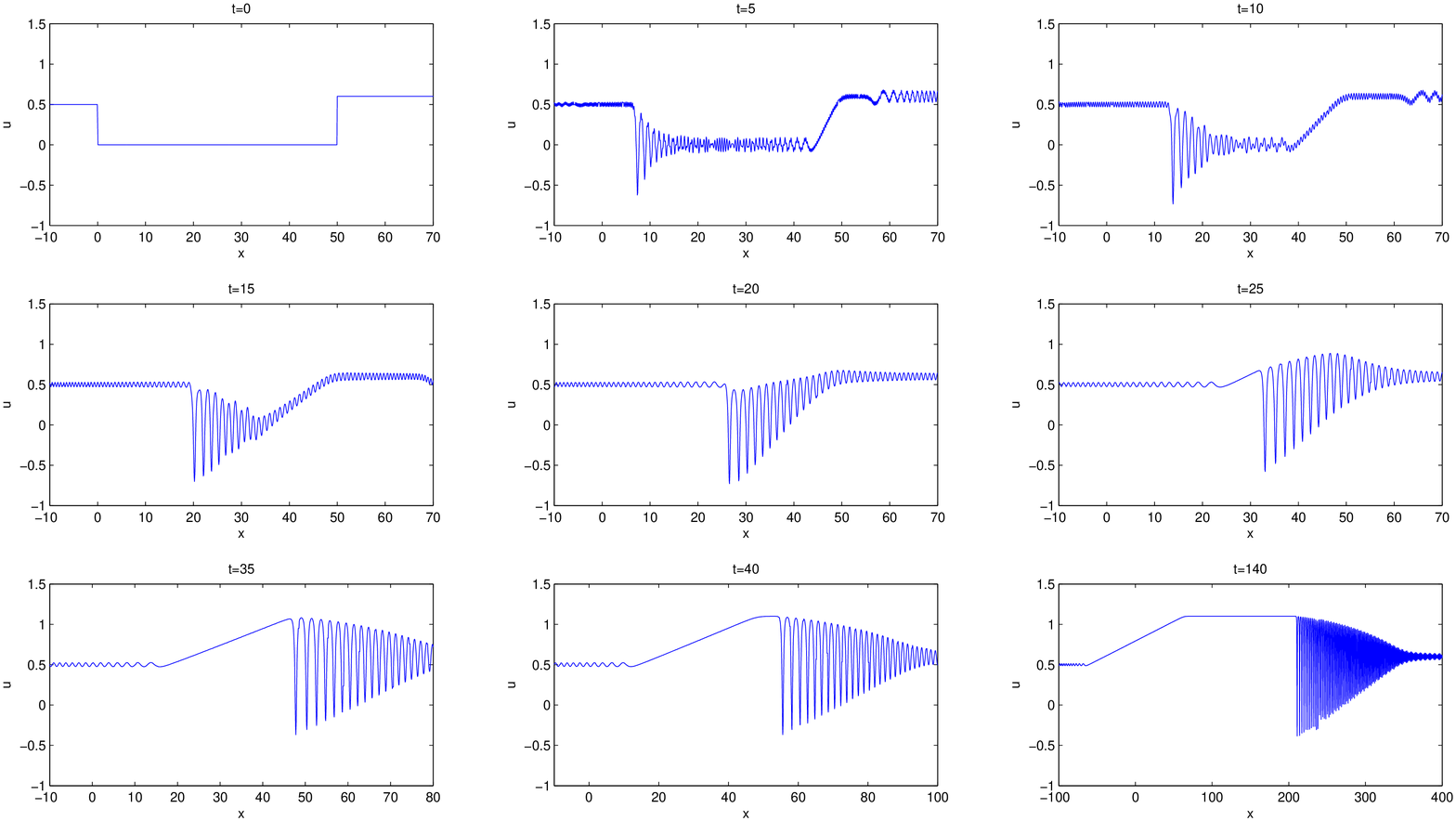}
\caption{Bidirectional interaction of a DSW and RW:  density (upper) and velocity (lower) profile; Initial data parameters: $A^+=1.5$, $A^- =-0.4 $, $l=50$.
The value of the dispersion parameter $\epsilon$ used in the simulations is $0.4$}
\label{fig3}
\end{center}
\end{figure}
Our concern  will be to obtain analytical description of the head-on DSW-RW interaction in terms of the initial profile parameters $A^+, A^-$ and $l$.

The evolution (\ref{NLS}), (\ref{ic2}) can be qualitatively understood using the results of papers
\cite{eggk95} and \cite{gk87}  where the  Riemann problem (which is a particular case of the problem (\ref{NLS}), (\ref{ic1})
 with $l=0$) was considered and a full classification for the different cases of the decay was constructed using similarity solutions of the modulation NLS-Whitham equations in the framework of the ``matched regularisation'' procedure of the Gurevich-Pitaevskii type (see also \cite{kod99}, \cite{bk06} for the further detailed analysis using the alternative ``global regularisation'' formulation).
  The crucial difference between the dispersive Riemann problem of \cite{eggk95},  \cite{gk87},  and the present problem (\ref{NLS}), (\ref{ic2}) is that the  two discontinuities for $\la_+$ and $\la_-$ are now spaced by a large distance $l$ so the modulation problem is no longer self-similar and a more general consideration is required.

The asymptotic solution of the NLS dispersive Riemann problem  obtained in \cite{gk87}, \cite{eggk95} (see also \cite{kod99}, \cite{bk06}) and our direct numerical simulations  of the more general general initial-value problem (\ref{NLS}), (\ref{ic2}) suggest that
the evolution  (\ref{NLS}), (\ref{ic2}) will initially lead to the formation of a right-propagating simple DSW and a left-propagating simple RW as in Figs.~1,2. Both waves expand with time and  start to
overlap and interact at some $t=t_0$. The interaction continues until some $t=t^*>t_0$  when the two waves fully separate
so that at $t>t^*$ there is a combination of new,  ``refracted'',  simple DSW and RW separated by a new constant state $n_0 \ne 1$, $u_0 \ne 0$. All the described stages of the DSW refraction are clearly seen on the direct numerical simulation plots in Fig.~3,~4.
\begin{figure}[h]
\begin{center}
\includegraphics[width=13cm]{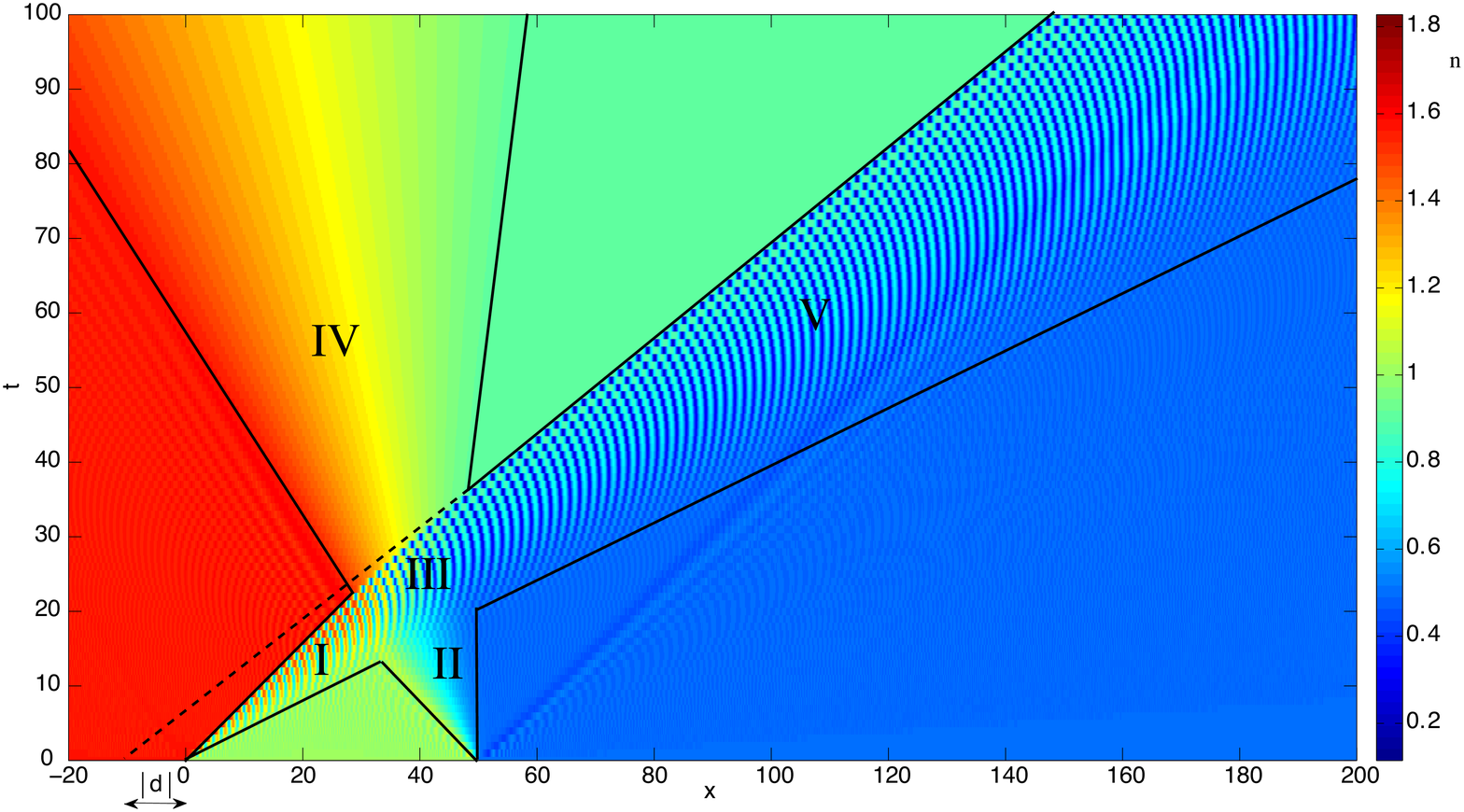}
\caption{(Colour online) Density plot  corresponding to the DSW-RW interaction shown in Fig.~3. The regions are as follows: I -- incident DSW; II -- incident RW; III -- DSW-RW interaction region;
 IV -- Refracted RW; V -- Refracted DSW. }
\end{center}\label{fig4}
\end{figure}
The most obvious effect of the head-on DSW-RW interaction  seen in Fig.~3 is the change of the key parameters (intensities, speeds) of the interacting waves. Another, more subtle, effect is the change of the phase distribution acquired by the DSW  during the interaction. The refraction phase shift $d$ of the DSW trailing dark soliton is shown in Fig.~4. Also we note that, if the incident DSW (RW) was centred at $t=0$, the refracted DSW (RW) will generally no longer be a centred wave. In Section 5 we shall construct an exact analytic solution of the NLS-Whitham equations asymptotically describing all stages of the head-on DSW-RW interaction seen in Figs.~3,~4 and compare the definitive DSW refraction parameters (the DSW amplification and acceleration coefficients as well as the refraction phase shift) with the corresponding parameters obtained numerically.

In conclusion of this section we note that the outlined head-on DSW-RW interaction can be naturally realised in the framework of the {\it dispersive piston problem}
(see \cite{hae08}, \cite{wing09}, \cite{kk10}) involving two pistons, the right piston being pulled out from the gas with constant velocity producing thus a left-propagating rarefaction wave while the left piston  being pushed into the gas producing the right propagating DSW. Another pertinent problem is the interaction of stationary {\it two-dimensional}  DSW and RW forming in
hypersonic dispersive flows past extended obstacles. This latter configuration  is relevant to the BEC experiments \cite{cornell05} and  can also be reformulated in terms of the already mentioned dispersive piston problem (see \cite{ek06}, \cite{wing09}).

\section{Refraction of shock waves in classical gas dynamics}

Before we proceed with the analysis of the bidirectional dispersive refraction problem (\ref{NLS}), (\ref{ic2}) we outline some
classical results on the head-on
interaction of viscous shocks and rarefaction waves (see e.g.  \cite{cf43}, \cite{moses}, \cite{RY}, \cite{ovs}).

Consider a one-dimensional motion of a polytropic isentropic gas, i.e. a gas with the equation of state $p=c n^\gamma$, where $p$ and $n$ are the gas pressure and density respectively,  $\gamma $ is the adiabatic exponent and $c$ is a constant
(the dispersionless shallow-water dynamics (\ref{eq21}) is equivalent to the dynamics of the polytropic gas with $\gamma = 2$).
 We consider the  following flow configuration (see Fig. 5). Let the gas motion at some moment of time, say $t=t_c \ge 0$ consist of  three regions of constant flow separated by two waves: a right-propagating shock wave (SW) located at some $x=x_c$ and a left-propagating RW centred at $x=l$ and occupying a finite region of space (as already was mentioned, such a configuration can be created by  piston motion inside a tube -- see e.g. \cite{moses}). Let the density and velocity of the flow  be
 $(n_1,u_1)$ as $x  \to  -\infty$ and $(n_2,u_2)$  as  $x  \to +\infty$.
 Then the gas motion at $t>t_c$ can be  qualitatively  described as follows:
 \begin{figure}[h]
\begin{center}
\includegraphics[width=6cm]{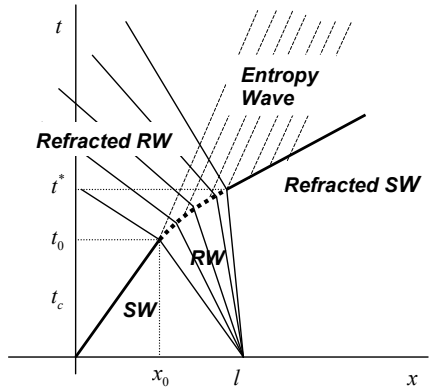}
\caption{Head-on interaction of SW and RW in classical gas dynamics}
\end{center}\label{fig4}
\end{figure}
\begin{itemize}
\item The SW and RW propagate independently until the moment $t=t_0$, when
the shock enters the rarefaction wave region at some $x=x_0$ say. Before that moment, i.e. for $0<t<t_0$, the entropy undergoes a rapid  {\it constant} change across the SW so the SW speed and strength (the  pressure excess across it) are determined by the standard Rankine-Hugoniot conditions. The RW is described by the centred left-propagating simple-wave solution of the inviscid hydrodynamic equations of motion. The  parameters of the constant flow between the SW and RW are found at the intersection of the $n$-$u$ diagrams for the SW and RW (see e.g. \cite{cf48}).
\item During certain time interval $t_0 < t< t^*$ the SW and the RW interact. The interaction is accompanied by the variations of the shock strength and results in the formation of the varying entropy region (the so-called `entropy wave')  behind the SW. Therefore, the flow behind the refracted SW is not isentropic.

    \item At $t=t^*$  the SW exits the RW region and the two waves again propagate separately in opposite directions, each having an altered (as compared with the values before the interaction) set of parameters.
       An important general result is that the speeds of the refracted SW and RW  and the density/velocity jumps across them are exactly as
    they would  have been in the corresponding origin-centred Riemann problem (i.e. in the decay of an initial discontinuity problem with the gas parameters $(n_1, u_1)$ at $x<0$ and $(n_2,u_2)$ at $x>0$ ), however, the spatial locations of the refracted waves differ from those in the corresponding Riemann problem. The refracted SW always has greater speed and strength than the original one.
\end{itemize}
As  already was mentioned, the presence of the  `entropy wave' behind the refracted SW radically complicates quantitative analysis of the motion and, as a result, the SW-RW head-on collision problem  can generally be treated only numerically (we stress that the notion of the entropy wave applies to viscous SWs in gas dynamics and is not applicable to, say, shallow water dynamics). In contrast to classical gas dynamics, dispersive-hydrodynamic flows governed by completely integrable equations often admit full analytical description. In particular, such a description is available for the DSW refraction process. This description can also  be  generalised (to some extent) to certain types of non-integrable dispersive equations.

\section{Single-phase modulation theory for the defocusing cubic NLS equation: account of results}
It is known very well that  analytical theory of one-dimensional dispersive compressible flows containing DSWs can be constructed  in the framework of the Whitham modulation equations \cite{whitham74}.
In this section we make a brief account
of the relevant results of the modulation theory for the defocusing
cubic NLS equation, which will be necessary for the analysis of the DSW - RW interaction in the subsequent sections. The single-phase  NLS modulation system was derived in \cite{fl86}, \cite{pavlov87} (see also \cite{kod99}) using the finite-gap integration methods.
A more elementary derivation of this system using a reduced version of the single-gap integration  can be found in
\cite{kamch2000}. Importantly, the theory presented in this section makes substantial use of the integrability of the NLS-Whitham modulation system, which is inherited from the complete integrability of the original cubic NLS equation.
A different method, proposed in \cite{el05} and applicable to the description of DSWs in nonintegrable systems,  will be used in Section 6 for the description of the DSW refraction in the media described by the NLS equation with saturable nonlinearity (\ref{snls}), which does not enjoy the complete integrability property.

It should be noted that, since the results of the
modulation theory do not depend on the value of the dispersion parameter $\epsilon$ in the NLS equation (\ref{3-1}), we
shall assume $\epsilon=1$ in the subsequent analytical representations of the periodic solutions, while in the numerical simulations we shall normally be using smaller values of $\epsilon$ to reduce the temporal scale of the DSW structure establishment.

\subsection{Periodic solution and modulation equations}

The periodic travelling wave solution of the defocusing NLS equation
(\ref{NLS})  can be expressed in terms of the Jacobi elliptic ${\rm sn}$ function
and is parametrised by four integrals of motion $\la_1\leq\la_2\leq\la_3\leq\la_4$ (as was already mentioned, we assume $\epsilon=1$ in the NLS equation),
\begin{equation}\label{eq013}
n =\frac14(\la_4-\la_3-\la_2+\la_1)^2+ (\la_4-\la_3)
(\la_2-\la_1)\,{\rm sn}^2\left(\sqrt{(\la_4-\la_2)(\la_3-\la_1)}\,
\theta,m\right) \, ,
\end{equation}
\begin{equation}\label{v}
u=U - \frac{C}{n} \, ,
\end{equation}
where $C=\frac{1}{8} (-\lambda_1 - \lambda_2 + \lambda_3 +
\lambda_4) (-\lambda_1 + \lambda_2 - \lambda_3 + \lambda_4)
(\lambda_1 - \lambda_2 - \lambda_3 + \lambda_4)$,
\begin{equation}\label{eq016}
\theta=x-Ut-\theta_0,\qquad U=\frac12 \sum_{i=1}^4\la_i,
\end{equation}
$U$ being the phase velocity of the nonlinear wave and $\theta_0$ the initial phase.

The modulus $0 \le m \le 1$  of the elliptic solution (\ref{eq013}) is defined as
\begin{equation}\label{eq015}
m=\frac{(\la_2-\la_1)(\la_4-\la_3)}{(\la_4-\la_2)(\la_3-\la_1)},
\end{equation}
and the wave amplitude is
\begin{equation}\label{amp}
a= (\la_4-\la_3)
(\la_2-\la_1) \, .
\end{equation}
The wavelength of the periodic wave (\ref{eq013}) is given by
\begin{equation}\label{eq017}
\begin{split}
\mathfrak{L}=  \int \limits_{\la_3}^{\la_4}
\frac{d\lambda}{\sqrt{(\la-\la_1)(\la-\la_2)(\la-\la_3)(\la_4-\la)}}=  \int \limits_{\la_1}^{\la_2}
\frac{d\lambda}{\sqrt{(\la-\la_1)(\la_2-\la)(\la_3-\la)(\la_4-\la)}} \\
=\frac{2{ \K}(m)}{\sqrt{(\la_4-\la_2)(\la_3-\la_1)}},
\end{split}
\end{equation}
${\K}(m)$ being the complete elliptic integral of the first kind. As a matter of fact, $\mathfrak{L}>0$.

In the limit   as $m \to 1$ (i.e. as $\la_3  \to \la_2$) the travelling wave
solution (\ref{eq013}) turns into a dark soliton
\begin{equation}\label{sol}
n=n_s - \frac{a_s}{\hbox{cosh}^2 (\sqrt{a_s}(x-U_st - \theta_0))}\, ,
\end{equation}
where the background density $n_s$, the soliton amplitude $a_s$  and
 velocity $U_s$  are expressed in terms of $\lambda_1, \lambda_2, \lambda_4$ as
\begin{equation}\label{22}
n_s=\frac{1}{4}(\la_4 - \la_1)^2,  \quad a_s=(\la_4 - \la_2)(\la_2 - \la_1) , \quad    U_s=\frac{1}{2}(\la_1+2\la_2+\la_4)\, .
\end{equation}
Allowing the parameters $\la_1, \la_2, \la_3, \la_4$ of the travelling wave solution (\ref{eq013}) to be slowly varying functions of $x$ and $t$,
one arrives, via the averaging or an equivalent multiple-scale perturbation procedure, at a  modulated nonlinear periodic wave
in which the evolution of ${\boldsymbol \la}=\{\la_1, \la_2, \la_3, \la_4 \}$
is governed by the Whitham  modulation equations \cite{fl86,pavlov87} (see \cite{whitham74,kamch2000}
for a detailed description of the Whitham method)
\begin{equation}
\label{eq18} \frac{\partial \la_i}{\partial
t}+V_i({\boldsymbol \la})\frac{\partial \la_i}{\partial x}=0, \qquad i=1,2,3,4,
\end{equation}
$\la_j$'s being the Riemann invariants.
The characteristic velocities  can be computed using the formula
\cite{gke92,kamch2000}
\begin{equation}
\label{eq019}
V_i(\boldsymbol{\la})=\left(1-\frac{\mathfrak{L}}{\partial_i\mathfrak{L}}\partial_i\right)U , \quad i=1,2,3,4 \, , \quad \hbox{where} \quad
\partial_i\equiv\partial/\partial \la_i \, .
\end{equation}
Substitution of Eq.~(\ref{eq017}) into Eq.~(\ref{eq019}) yields the
explicit expressions
\begin{equation}\label{vi}
\begin{split}
V_1&=\tfrac12 \sum \la_i
-\frac{(\la_4-\la_1)(\la_2-\la_1)}{(\la_4-\la_1)-(\la_4-\la_2)\mu(m)},\\
V_2&=\tfrac12 \sum \la_i
+\frac{(\la_3-\la_2)(\la_2-\la_1)}{(\la_3-\la_2)-(\la_3-\la_1)\mu(m)},\\
V_3&=\tfrac12 \sum \la_i
-\frac{(\la_4-\la_3)(\la_3-\la_2)}{(\la_3-\la_2)-(\la_4-\la_2)\mu(m)},\\
V_4&=\tfrac12 \sum \la_i
+\frac{(\la_4-\la_3)(\la_4-\la_1)}{(\la_4-\la_1)-(\la_3-\la_1)\mu(m)},
\end{split}
\end{equation}
where  $\mu(m)=\E(m)/ \K(m)$, $\E(m)$ being the complete elliptic integral
of the second kind.
The characteristic velocities  (\ref{vi})  are real for all values of the Riemann
invariants, therefore system (\ref{eq18})
is hyperbolic. Moreover, it is not difficult to show using representation (\ref{eq019}) that
\begin{equation}\label{gn}
\partial_i V_i >0\, \quad \hbox{for all} \quad i\, ,
\end{equation}
so the NLS-Whitham system (\ref{eq18}), (\ref{vi})
is {\it genuinely nonlinear} \cite{lax}. Indeed, differentiating
(\ref{eq019}) we get:
\begin{equation}\label{e}
\partial_i V_i = \frac{\mathfrak{L} }{2(\partial _i\mathfrak{L})^2}\partial ^2_{ii}\mathfrak{L} \, .
\end{equation}
Using the integral representations (\ref{eq017})  for $\mathfrak{L}$ one can readily see that $\partial ^2_{ii}\mathfrak{L}>0$ for all $i$ (it is convenient to use the first representation for the differentiations with respect $\la_1$ and $\la_2$ and the second one for the differentiations with respect $\la_3$ and $\la_4$), which immediately implies
(\ref{gn}).

Also, using (\ref{eq019}) and the intergral representations (\ref{eq017}) one can readily show by a direct calculation that
\begin{equation}\label{order}
i>j \quad \hbox{implies} \quad   V_i>V_j \, .
\end{equation}
Thus, the ordering $\la_1\leq\la_2\leq\la_3\leq\la_4$ of the Riemann invariants implies a similar ordering $V_1 \le V_2 \le V_3 \le V_4$ for the characteristic velocities.
We note that the properties (\ref{gn}) and (\ref{order}) were established in \cite{jl99}, \cite{kod99} using the finite-gap integration framework for the derivation of the Whitham equations.

For the DSW analysis in the subsequent sections we shall need the reductions of formulae (\ref{vi})
for the limiting cases $m=0$ (harmonic limit) and $m=1$ (soliton limit).

The harmonic limit $m=0$ can be achieved in one of the two possible ways:
either via $\la_2=\la_1$ or via $\la_3=\la_4$.
Then:
\begin{eqnarray}
\hbox{when} \ \  \la_2=\la_1:   \quad    V_2=V_1=\lambda_1 + \frac{\lambda_3 + \lambda_4}{2} +
\frac{2(\lambda_3-\lambda_1)(\lambda_4 - \lambda_1)}{2\lambda_1 - \lambda_3-\lambda_4} \, ,&&
\nonumber \\
&& \label{m01} \\
V_3= \frac{3}{2} \la_{3} + \frac12{\la_{4}}=V_-(\la_3, \la_4)\, , \quad
V_4= \frac{3}{2} \la_{4} + \frac12{\la_{3}} = V_+(\la_3, \la_4)\, .&& \nonumber
\end{eqnarray}
\begin{eqnarray}
\hbox{when} \ \  \la_3=\la_4:   \quad    V_3=V_4=\lambda_4 + \frac{\lambda_1 + \lambda_2}{2} +
\frac{2(\lambda_4-\lambda_2)(\lambda_4 - \lambda_1)}{2\lambda_4 - \lambda_2-\lambda_1} \, ,&&
\nonumber \\
&& \label{m02} \\
V_1= \frac{3}{2} \la_{1} + \frac12{\la_{2}}=V_-(\la_1, \la_2)\, , \quad V_2= \frac{3}{2} \la_{2} + \frac12{\la_{1}}=V_+(\la_1, \la_2)\, . && \nonumber
\end{eqnarray}
In the soliton limit we have $m=1$.
This can happen only if $\la_2=\la_3$, so we obtain:
\begin{eqnarray}
\hbox{when} \ \  \la_2=\la_3:   \quad    V_2=V_3=\frac{1}{2}(\lambda_1+ 2\lambda_2 + \lambda_4)=U_s\, ,&&
\nonumber \\
&& \label{m1} \\
V_1= \frac{3}{2} \la_{1} + \frac12{\la_{4}}=V_-(\la_1, \la_4)\, , \quad V_4= \frac{3}{2} \la_{4} + \frac12{\la_{1}}=V_+(\la_1, \la_4)\, . && \nonumber
\end{eqnarray}

Thus, in both  harmonic  and soliton    limits
the fourth-order modulation system (\ref{eq18}),
(\ref{vi}) reduces to the system of three equations, two of which
are decoupled.  Moreover, one can see that in all considered limiting cases
the decoupled equations agree with the {\it dispersionless limit} of the NLS
equation (\ref{3-1}). This property makes possible the matching of the modulation solution
with the solution to the dispersionless limit equations at the  points where $m=0$ or $m=1$.

\subsection{Free-boundary matching conditions for the modulation equations}

In the description of a DSW,  the Whitham
equations (\ref{eq18}) must be endowed with certain initial or boundary
conditions for the Riemann invariants $\lambda_i$. We shall be using the Gurevich-Pitaevskii
type boundary-value (matching) problem first formulated in \cite{gp74} for the KdV dispersive shock waves and extended to the NLS
case in \cite{gk87}. A different  type of the problem formulation (the so-called regularised initial-value problem for the NLS-Whitham equations) proposed in  \cite{kod99} and recently used in \cite{bk06}, \cite{hec09} for the  {\it numerical analysis} of the DSW interaction  is less convenient for our purposes as the {\it analytical description} of the interaction zone requires the hodograph solutions of the Whitham equations, and the poor compatibility of the initial-value problems with the hodograph transform is well known  (see e.g. \cite{whitham74}). The Gurevich-Pitaevskii matching conditions, on the contrary, are ideally compatible with the hodograph transform as they turn into the classical Goursat type characteristic boundary conditions on the hodograph plane (see \cite{gke91}, \cite{gke92}, \cite{ek95}, \cite{wing09}).
It is clear that both formulations (regularised initial-value problem for the Whitham equations and the Gurevich-Pitaevskii type matching problem) must be equivalent, although we are not aware of the rigorous proof of this equivalence.

To be specific,  we shall formulate  boundary (matching) conditions for
the right-propagating DSW.   Without loss of generality
we assume that the formation of the
DSW starts at the origin of the $(x,t)$-plane. In the Gurevich-Pitaevskii
setting, the upper $(x,t)$-half plane is split into three regions (see Fig.~6):
$(-\infty, x^-(t))$, $[x^-(t), x^+(t)]$ and $(x^+(t), + \infty)$.
\begin{figure}[ht]
\centerline{\includegraphics[width=8cm]{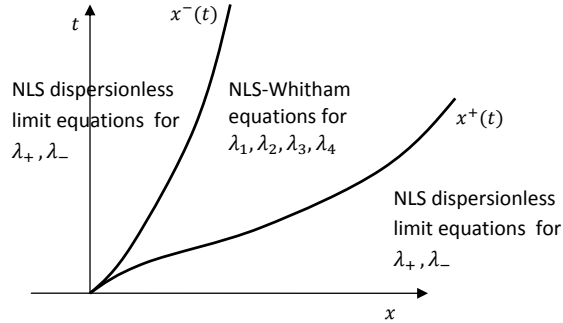}}
\caption{
Splitting of the $xt$-plane in the Gurevich-Pitaevskii problem for the
defocusing NLS equation.
}
\label{fig7}
\end{figure}

In the ``outer'' regions $(-\infty, x^-(t))$ and $(x^+(t), + \infty)$ the flow is
governed by the
dispersionless limit of the NLS equation, i.e. by the shallow-water system
(\ref{er}), (\ref{V}) for the Riemann invariants $\la_{\pm}$. In the DSW region
$[x^-(t), x^+(t)]$ the averaged oscillatory flow is described by four Whitham
equations (\ref{eq18}) for the Riemann invariants $\lambda_j$
with the following matching conditions at the trailing $x^-(t)$ and leading
$x^+(t)$ edges of the DSW
(see \cite{gk87,ek95} for details):
\begin{equation}
\begin{array}{l}
\hbox{At} \ \ x=x^-(t):\qquad \la_3=\la_2\, , \  \ \la_4 = \la_+ , \  \   \la_1 = \la_- \, , \\
\hbox{At} \ \ x=x^+(t):\qquad \la_3=\la_4\, , \ \  \la_2 = \la_+ ,  \  \  \la_1 = \la_- \, .
\end{array}
\label{bc}
\end{equation}
Here $\la_{\pm}(x,t)$ are the Riemann invariants of the dispersionless limit of
the NLS equation in the
hydrodynamic form (\ref{er}), (\ref{V}).
The free boundaries $x^{\pm}(t)$ of the DSW are defined by the kinematic conditions
\begin{equation}\label{mult}
\frac{dx^-}{dt}=V_ 2(\la_1, \la_2, \la_2, \la_4)=V_3(\la_1, \la_2, \la_2, \la_4)\, ,
\qquad \frac{dx^+}{dt}=V_3(\la_1, \la_2, \la_4, \la_4)=V_4(\la_1, \la_2, \la_4,  \la_4)
\end{equation}
and so are   multiple characteristics of the Whitham system. The multiple
characteristic velocities $V_2=V_3$  and
$V_3=V_4$ in (\ref{mult}) are explicitly given  by equations (\ref{m1}) and  (\ref{m02})
respectively. One should stress that determination of $x^{\pm}(t)$
is an inherent part of the construction of the full modulation solution.
We also emphasize that matching conditions (\ref{bc})
are consistent with the  structure of the Whitham system
(\ref{eq18}), (\ref{vi})  in the limiting cases $m=0$ and $m=1$  (see (\ref{m02}), (\ref{m1}))
and with the spatial oscillatory profile of the DSW
in the defocusing NLS dispersive hydrodynamics (as is known very well, such a DSW
has a dark soliton $(m=1)$ at the trailing edge and degenerates into the
vanishing amplitude
harmonic wave ($m=0$) at  the leading edge---see
\cite{gk87,eggk95,kgk04,ha06}).

\subsection{Hodograph transform and the mapping to the Euler-Poisson-Darboux equation}

The hydrodynamic type modulation system (\ref{eq18}), (\ref{vi}) can be reduced to
a system of linear partial differential equations using the (generalised) hodograph transform \cite{ts85}.
We first fix two of the Riemann invariants, say
\begin{equation}\label{}
\la_i=\la_{i0}=\hbox{constant}\, , \quad
\la_j=\la_{j0}=\hbox{constant}\, , \qquad i \ne j \, ,
\end{equation}
to reduce (\ref{eq18}) to the system of two equations for the remaining two invariants
$\la_k(x,t)$ and $\la_l(x,t)$,  $k \ne l \ne i \ne j$
\begin{equation}\label{34}
\frac{\partial \la_k}{\partial t}+V_k(\la_k, \la_l)\frac{\partial
\la_k}{\partial x}=0\, , \qquad \frac{\partial \la_l}{\partial
t}+V_l(\la_k, \la_l)\frac{\partial \la_l}{\partial x}=0\, ,
\end{equation}
where $V_{k,l}(\la_k, \la_l)  \equiv  V_{k,l}(\la_{i0}, \la_{j0}, \la_k, \la_l)$.
Applying the hodograph transform to system (\ref{34}) we arrive at
a linear system for $x(\la_k, \la_l)$, $t(\la_k, \la_l)$,
\begin{equation}\label{hod1}
\frac{\partial x}{\partial \la_k}-V_l(\la_k, \la_l)\frac{\partial t}{\partial
\la_k}=0\, , \qquad \frac{\partial x}{\partial
\la_l}-V_k(\la_k, \la_l)\frac{\partial t}{\partial \la_l}=0 \, .
\end{equation}
Note that the hodograph transform requires that $\partial_x \la_{k,l} \ne 0$.
Now we make in (\ref{hod1}) the change of variables
\begin{equation}\label{Ts1}
 x - V_k t = W_k  \, , \qquad   x - V_l t = W_l   \, ,
\end{equation}
which reduces it to a symmetric system for $W_k(\la_k,\la_l)$, $W_l(\la_k,\la_l)$:
\begin{equation}\label{Ts2}
 \frac{\partial _k W _l}{W_k - W_l} = \frac{\partial_k
V_l}{V_k - V_l} \ ;  \quad  k \ne l ; \quad
\partial_k\equiv\partial/\partial \la_k \, .
\end{equation}
The symmetry between $V_l$ and $W_l$ in (\ref{Ts2}) and the `potential' structure (\ref{eq019}) of the
vector function $(V_{k}, V_l)$ implies the possibility of introducing
a single scalar function $g(\la_k, \la_l)$  instead of the vector  $(W_{k}, W_l)$:
\begin{equation}
\label{scalar}
W_i=\left(1-\frac{\mathfrak{L}}{\partial_i\mathfrak{L}}\partial_i\right)g \, ,\quad
i=k,l,
\end{equation}
or, which is the same (use (\ref{eq019})),
\begin{equation}\label{scalar1}
W_i=g+ 2(V_i-U) \partial_i g \, , \quad
i=k,l\, .
\end{equation}
Then substituting (\ref{eq019}), (\ref{scalar}) into (\ref{Ts2}) we arrive, taking into account (\ref{eq017}),
at the Euler-Poisson-Darboux (EPD) equation for $g(\la_k, \la_l)$,
\begin{equation}\label{EPD}
2(\la_l-\la_k)\partial^2_{kl} g
=\partial_l g - \partial_k g \, .
\end{equation}
The EPD equation was first derived in the present NLS context in \cite{gke92} and  was later used in \cite{ek95} and \cite{tianNLS} for the construction of the general solution of the semi-classical defocusing NLS equation with smooth monotonically decreasing initial data.

Note that system (\ref{34}) essentially describes the interaction of two simple waves of modulation of the NLS equation
so the fact that this system reduces, in the hodograph plane, to the EPD equation (\ref{EPD}) describing interaction of two simple waves in classical dispersionless
shallow-water theory (or in gas dynamics of polytropic isentropic gas with $\gamma = 2 $) (see \cite{whitham74} for instance) is quite remarkable.

The general solution of the EPD equation (\ref{EPD}) can be represented in
the form (see, for instance, \cite{tricomi})
\begin{equation}\label{gs}
g=\int \limits _{a_1} ^{\la_k} \frac{\phi_1(\la) d
\la}{\sqrt{(\la - \la_k)(\la_l-\la)}}
+
\int \limits _{a_2} ^{\la_l} \frac{\phi_2(\la) d \la}{\sqrt{(\la- \la_k)(\la_l-\la)}} \, ,
\end{equation} where $\phi_{1,2}(\la)$ are arbitrary (generally,
complex-valued) functions and $a_{1,2}$ are arbitrary constants (which could be absorbed into $\phi_{1,2}$).

As a matter of fact, the same construction can be realized for any
pair of  Riemann invariants while the two remaining invariants
are fixed. Moreover, equations (\ref{Ts1}) -- (\ref{Ts2}) and further (\ref{eq019}) -- (\ref{EPD}) turn out
to be valid even when all four Riemann invariants vary
\cite{gke92,ek95}.
This becomes possible for two reasons. Firstly, the
NLS modulation system (\ref{eq18}), (\ref{eq019}) is integrable via
the generalized hodograph transform \cite{ts85} which reduces it
to overdetermined consistent system (\ref{Ts2}), where
$k,l=1,2,3,4$, $k \ne l$. Secondly, the ``potential'' structure of the
characteristic speeds (\ref{eq019}) makes it possible to use the same
substitution (\ref{scalar}) for all $k=1,2,3,4$ which results in the
 consistent system of six EPD equations (\ref{EPD})
involving all pairs $\lambda_k, \lambda_l$, $k \ne l$.

Thus, the problem of  integration of the nonlinear Whitham system (\ref{eq18})
with rather complicated coefficients (\ref{vi}) is essentially reduced to solving the
classical linear EPD equation
(\ref{EPD}). Essentially, one needs to express the functions $\phi_{1,2}(\la)$
in the general solution (\ref{gs}) in terms of the initial or boundary conditions
for the NLS equation (\ref{3-1}). As was shown in \cite{gke92}, \cite{ek95} (see also \cite{wing09})
the free-boundary  nonlinear matching conditions (\ref{bc})  are most conveniently
translated into a classical linear Goursat characteristic boundary problem for the
EPD equation (\ref{EPD}). This enables one to find the unknown functions $\phi_{1,2}(\la)$ in terms of
Abel transforms of the initial data.

In conclusion of this section we note that hodograph solutions do not include the special
family of the simple-wave solutions as the latter correspond to the vanishing
Jacobian
of the hodograph transform $(\la_k, \la_l) \mapsto (x,t)$ (see, for instance,
\cite{whitham74}).

%Still, the similarity $(x/t)$ modulation solution can be formally obtained from the hodograph solution represented in the symmetric %form (\ref{Ts1}). For instance,  putting
%$W_3=0$ and setting  all the Riemann invariants $\la_j$ with
%$j \ne 3$ constant one arrives at the  similarity modulation solution, in which $\la_3=\la_3(x/t)$
%is implicitly specified by the equation $V_3=x/t$.

\subsection{Modulation phase shift}
In the modulated wave, the  initial phase $\theta_0$ of the periodic solution  (\ref{eq013}) -- (\ref{eq015}) is no longer an independent constant parameter but  rather  a slow function of $x,t$ so it is better described as the modulation phase shift. As was shown in \cite{kpt08}, the function  $\theta_0(x,t)$ can be found from the requirement that the local wavenumber $k=2\pi/\mathfrak{L}$ and
the local frequency $\omega=kU$ in the modulated wave (\ref{eq013})  must satisfy the generalised phase relationships
\begin{equation}\label{Phase}
k = \Theta_x\, , \qquad  \omega= - \Theta_t\, ,
\end{equation}
where
\begin{equation}\label{Theta}
\Theta=k \theta= k x- \omega t- k \theta_0 \,
\end{equation}
is the angular phase. Relationships (\ref{Phase}) imply the `conservation of waves' law
\begin{equation}\label{}
k_t+\omega_x=0 \, ,
\end{equation}
which is consistent with the modulation system (\ref{eq18}) and thus yields the representation $V_i=\partial_i \omega/\partial_i k$
for the characteristic speeds, equivalent to (\ref{eq019}).

For the general modulation relationship (\ref{Phase}) to be consistent with the linear $x,t$-dependence of the phase (\ref{Theta}) entering the local single-phase NLS solutions (\ref{eq013}), (\ref{eq015}) one must  assume $\theta_0(x,t) = \vartheta_0(\la_1,\la_2, \la_3, \la_4)$, which implies that the phase shift is completely determined by the evolution of the Riemann invariants $\la_j(x,t)$ in the modulation solution.
To find $\vartheta_0(\la_1,\la_2, \la_3, \la_4)$ we differentiate
(\ref{Theta}) with respect to $x$ to obtain
\begin{equation}\label{thex}
\Theta_x=k + \sum \limits_{i=1}^4 \left\{ x \partial_i k -  t \partial_i \omega  - \vartheta_0 \partial_i k  -k
\partial_i \vartheta_0\right \} \partial_x \la_i \, .
\end{equation}
Comparing (\ref{thex}) with  (\ref{Phase}) we obtain for any  pair $i,j$, $i \ne j$
\begin{equation}\label{44}
x \partial_i k -  t \partial_i \omega  - \vartheta_0 \partial_i k  -k \partial_i \vartheta_0 =0 \, , \qquad
x \partial_j k -  t \partial_j \omega  - \vartheta_0 \partial_j k  -k \partial_j \vartheta_0 =0 \, ,
\end{equation}
provided $\partial_x \la_{i,j} \ne 0$. On using $\partial_i \omega/\partial_i k = V_i$  and $k=2\pi / \mathfrak{L}$ system (\ref{44})
is readily transformed to the form
\begin{equation}\label{hodtheta}
x-V_nt= \left(1-\frac{\mathfrak{L}}{\partial_n\mathfrak{L}}\partial_n\right) \vartheta_0 \,  , \qquad n=i,j, \quad i \ne j\, .
\end{equation}
Comparison of expression (\ref{hodtheta}) with the modulation hodograph solution (\ref{Ts1}), (\ref{scalar}) enables one to identify the modulation phase shift $\vartheta_0(\boldsymbol{\la})=\theta_0(x,t)$
with the solution $g(\boldsymbol{\la})$ to the relevant boundary value problem for the EPD equation (\ref{EPD}), i.e.
\begin{equation}\label{sh}
\theta_0(x,t)=g(\boldsymbol{\la}(x,t)) \, .
\end{equation}

One can also see from (\ref{hodtheta}) that
one should set $\theta_0=0$ for a simple centred DSW described by the modulation solution in which all but one Riemann invariants
are  constants and the varying invariant, say $\la_m$, is implicitly specified by the equation  $x-V_m t=0$. The condition $\theta_0=0$ then implies that in the dispersive Riemann (decay of a step) problem the DSW trailing dark soliton  (\ref{sol}) is centred exactly at the trailing edge $x^-(t)$ defined by   (\ref{mult}), (\ref{m1}).

\section{Interaction of  DSW and RW: modulation solution}
\subsection{Before interaction, $0<t<t_0$}
At $t=0$,  a simple origin-centred right-propagating DSW is generated due to the jump of the Riemann invariant $\la_+$ while the jump of $\la_-$
produces a similarity ``shallow-water'' rarefaction wave centred at $x=l$ and propagating to the left.
\begin{figure}[ht]
\begin{center}
\includegraphics[width=8cm]{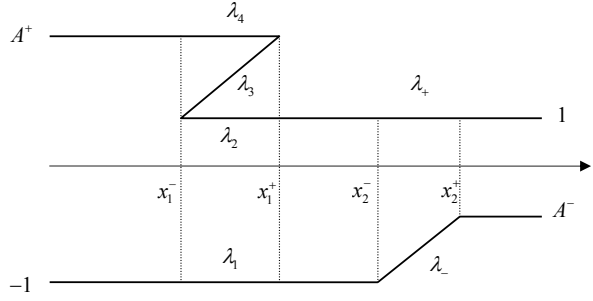}
\caption{
Schematic behaviour for the Riemann invariants before the interaction of the DSW and RW,  $0<t<t_0$.
}
\end{center}
\end{figure}

The similarity modulation solution describing the DSW has the form  \cite{gk87}, \cite{eggk95}
\begin{equation}\label{DSW1}
\begin{split}
\la_1=-1\, ,\quad \la_2=1\, , \quad \la_4=A^+\, , \\
\frac{x}{t}=V_3(-1,1,\la_3, A^+)=\frac{\la_3+A^+}{2} - \frac{(A^+-\la_3)(\la_3-1)}{\la_3-1 - (A^+ - 1)\mu(m)}\, ,
\end{split}
\end{equation}
where
\begin{equation}\label{msim}
m=\frac{2(A^+ - \la_3)}{(A^+ - 1)(\la_3+1)}\, .
\end{equation}
The boundaries of the DSW are then found from (\ref{DSW1}) by setting $\la_3=A^+$ (i.e. $m=0$)
for the leading edge $x^+_1$ and $\la_3=1$ (i.e. $m=1$) for the trailing edge $x^-_1$:
\begin{equation}\label{bound1}
x^-_1=\frac{1+A^+}{2} t\, , \qquad x^+_1=(2A^+ - \frac{1}{A^+})t\, .
\end{equation}
The dark soliton at the trailing edge $x^-_1$ of the DSW has the amplitude $a_s$ and rides on the background $n_s$ defined by
(see (\ref{22}))
\begin{equation}\label{solamp1}
n_s=(1+A^+)^2/4 \, , \quad a_s=2(A^+-1)\, .
\end{equation}
The value $A^+=3$ corresponds to the formation of a  vacuum point  at the trailing edge of the DSW \cite{eggk95} so that the density at the
dark soliton minimum is $n_s-a_s=0$. For $A^+>3$ the vacuum point occurs inside the DSW at some $x=x_v$, where $x^-<x_v<x^+$ --- see details in \cite{eggk95}, \cite{ha06}.

We  define the  {\it relative} intensity (hereafter -- simply intensity) $I$  of a DSW as the  density ratio  across it:
\begin{equation}\label{int}
I=\frac{n_1}{n_2}\, ,
\end{equation}
where $n_1$ and $n_2$ are the values of density upstream and downstream the DSW respectively. This definition can be related to the one accepted in classical gas dynamics, where the relative  pressure excess across the SW is often used as a measure of the SW strength. One should, however, stress that the notion of the DSW intensity for the NLS flows retains its original meaning only
for DSWs not containing vacuum points. The modification of the flow resulting from the vacuum point appearance will be discussed below in Section 5.3.3.

For the  incident DSW (i.e. before the interaction) we obviously have $n_1=n_s$ and $n_2=1$, i.e.
its intensity is
\begin{equation}\label{i0}
I_0=\frac{(1+A^+)^2}{4} \, .
\end{equation}
Now we turn to the left-propagating rarefaction wave, which is asymptotically described by the centred at $x=l$ similarity solution of the classical
limit equations (\ref{er}), (\ref{V}) (see, e.g., \cite{cf48}, \cite{ll}) :
\begin{equation} \label{rar1}
\la_+=1 \, ,
\end{equation}
 \begin{eqnarray}
\   \  \ \la_-&=& -1, \qquad  \qquad \qquad x < x^-_2 ;\nonumber \\
\frac{x-l}{t}=V_-(\la_-,1)&=&\frac{3\la_-+1}{2}\, , \    \ \qquad x^-_2 \le x \le x^+_2; \label{rar11}\\
 \   \  \ \la_{-}&=& A^-\, , \qquad  \qquad \qquad x>x^+_2 . \nonumber
\end{eqnarray}
Here the boundaries $x^{\pm}_2$ are specified by the formulae:
\begin{equation}\label{bound2}
x^-_2=l-t\, , \qquad x_2^+=l+\frac{3A^-+1}{2}t \, .
\end{equation}
It is instructive to note that, since the modulation system (\ref{eq18}) in the harmonic limit  is consistent with the shallow-water equations (\ref{er}), (\ref{eq20})  --- see (\ref{m02}),
 the RW solution (\ref{rar1}), (\ref{rar11}) is also a solution of full modulation system (\ref{eq18}), namely
\begin{equation}\label{restr1}
  \la_3=\la_4 = A^+; \qquad \la_2= \la_+=1 \, , \qquad \la_1=\la_-(x,t) \, .
\end{equation}
This identification of the RW solution of the dispersionless limit equations as a particular  solution of the full modulation system will be used in Section 5.2.
The schematic behaviour of the Riemann invariants during the first stage of evolution, before the DSW-RW interaction, is shown in Fig.~7.

One can readily see that $d x_1^+/dt > d x_2^-/dt$ (this also  follows from the characteristic velocity ordering described in Section 4.1) so the DSW will start overtaking the RW at some  moment $t=t_0$ when the leading edge of the DSW will catch up the
trailing edge of the RW at  $x_0=x^+_1(t_0)=x^-_2(t_0)$.
Using (\ref{bound1})  and (\ref{bound2}) we obtain
\begin{equation}\label{}
t_0=\frac{A^+l}{2(A^+)^2+A^+-1}\, , \qquad x_0=\frac{2(A^+)^2-1}{2(A^+)^2+A^+-1}l\, .
\end{equation}
\subsection{Interaction, $t_0<t<t^*$}
At $t=t_0$ the DSW enters the  RW region so that at $t>t_0$ a nonlinear interaction zone  confined to the interval $[x_2^-, x_1^+]$ forms (see Fig.~8) and evolves until some moment
$t^*$ when the DSW completely overtakes the RW so that $x_2^+(t^*) =  x_1^-(t^*)$.  At $t>t^*$ the DSW and RW fully separate, each acquiring a new set of parameters $\la_j$ compared
to their initial characterization. One should stress that, for $t>t_0$
the functions $x_1^\pm(t)$ and $x_2^\pm(t)$ are no longer described by the formulae (\ref{bound1}), (\ref{bound2}) from the previous subsection.

\begin{figure}[ht]
\begin{center}
\includegraphics[width=8cm]{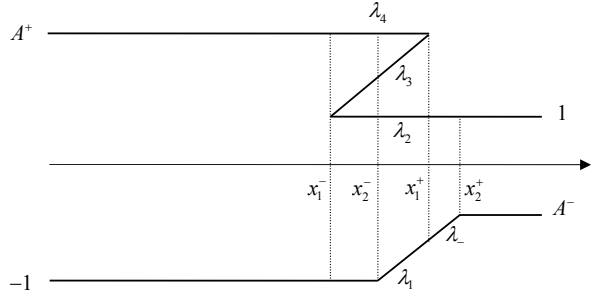}
\caption{
Schematic behaviour of the Riemann invariants during interaction of the DSW and RW, $t_0<t<t_*$.
}
\end{center}
\end{figure}
The corresponding interaction diagram in the $(x,t)$-plane is shown in Fig.~9 (left). One can see that the NLS DSW-RW interaction
in the semiclassical limit is essentially described by the interaction of two rarefaction fans: one of the shallow-water equations and another one --- of the Whitham equations.

In the interaction region $[x_2^-, x_1^+]$ one still has $\la_2=1$ and $\la_4=A^+$ but the remaining
two Riemann invariants ($\la_1$ and $\la_3$) now vary so the modulation solution is no longer self-similar and
a more general, hodograph solution (\ref{Ts1}) is needed.
This is found via  additional transformation
(\ref{scalar}) reducing Tsarev's equations (\ref{Ts2}) for $W_{1,3}(\la_1, \la_3) \equiv W_{1,3}(\la_1, 1, \la_3, A^+)$ to the EPD equation (\ref{EPD}).
The general solution (\ref{gs}) of the EPD equation is parametrised by two arbitrary functions
$\phi_{1,2}(\la)$ which should be found from appropriate boundary conditions. These conditions, in their turn, must
follow from the continuity matching conditions for $\la_1$ and $\la_3$ at the unknown boundaries
$x_2^-(t)$ and $x_1^+(t)$.
\begin{figure}[ht]
\begin{center}
\includegraphics[width=8cm]{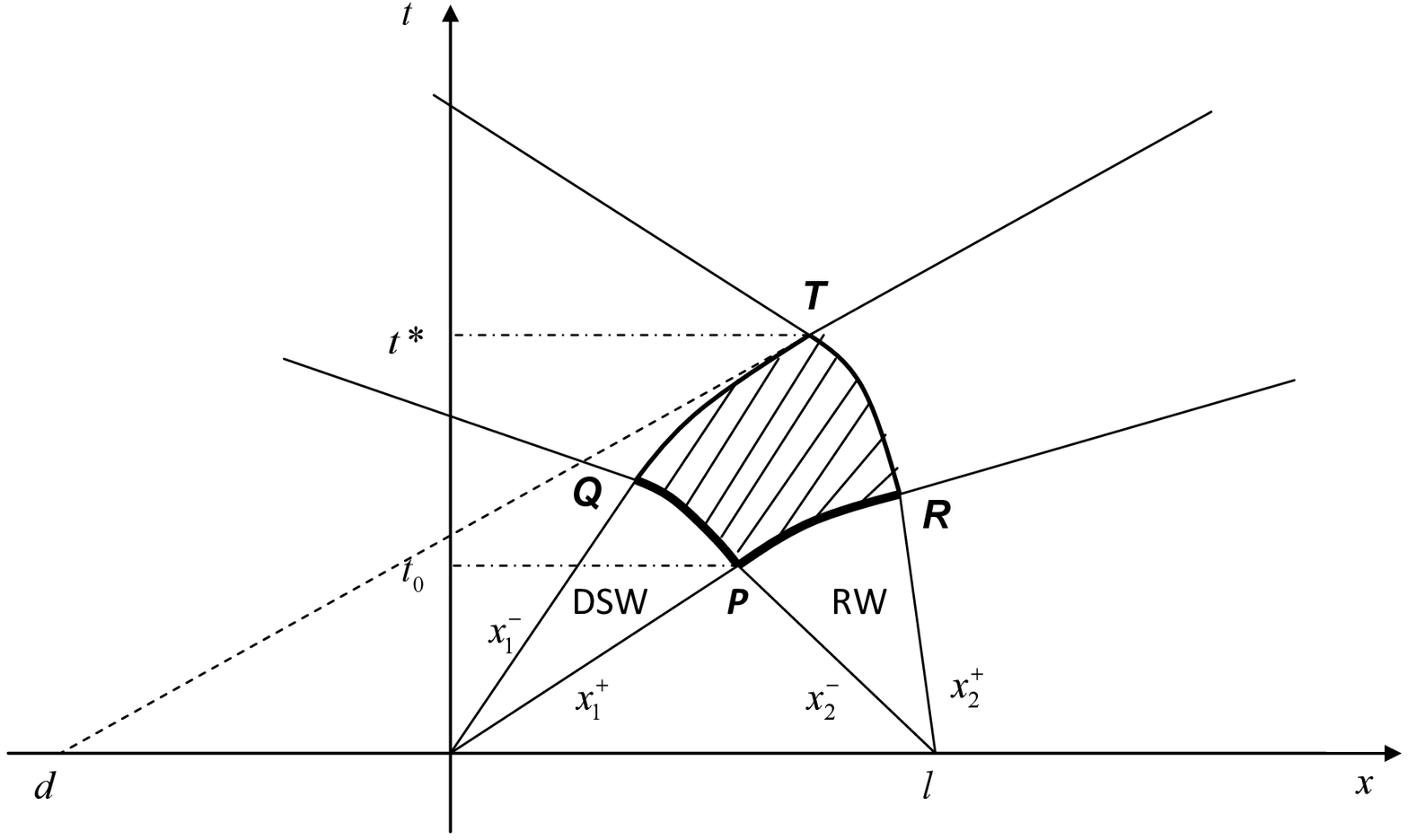} \qquad \quad \includegraphics[width=4 cm]{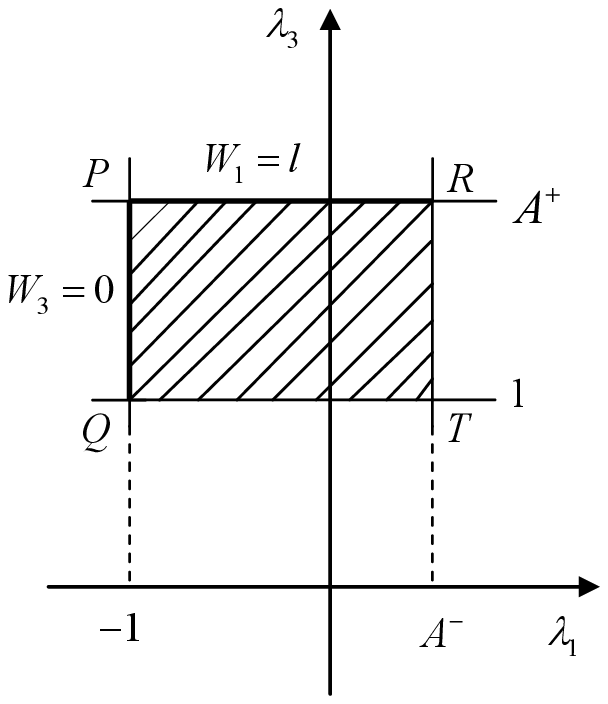}
\caption{DSW-RW interaction diagram. Left: physical, $(x t)$ plane; Right: hodograph, $(\la_1 \la_3)$ plane.
}
\end{center}
\end{figure}

At the left boundary $x=x^-_2(t)$ of the interaction zone (segment $PQ$ in the interaction diagram in Fig. 9, left) we have
\begin{equation}\label{left1}
 \la_1=-1\, ,  \quad  \la_2 = 1 , \  \  \la_3=\la_3^{s}(x^-_2(t),t) \, , \  \  \la_4 = A^+ \, , \\
\end{equation}
where $\la_3^s(x,t)=\la_3(x/t)$  is found from the similarity modulation solution (\ref{DSW1}).
At the right boundary   $x=x^+_1(t)$ of the interaction zone (segment $PR$ in Fig.~9 left)
we have, similar to the second condition (\ref{bc}),
\begin{equation}\label{right1}
\la_1 = \la^r_-(x^+_1(t),t)\, , \ \  \ \la_2 = 1 ,  \  \ \la_3= \la_4=A^+  \, ,
\end{equation}
 and $\la^r_-(x,t)=\la_-(\frac{x-l}{t})$ is found from the rarefaction wave solution (\ref{rar1}).

We now need to translate  nonlinear free-boundary conditions (\ref{left1}) and (\ref{right1}) into the boundary
conditions for the function $g(\la_1, \la_3)$ satisfying  the EPD equation
\begin{equation}\label{EPD1}
2(\la_3-\la_1)\partial^2_{13} g
=\partial_3 g - \partial_1 g \, , \qquad \partial_j \equiv \partial/\partial \la_j \, .
\end{equation}
This is done in two steps. First we derive the boundary conditions for the functions $W_1(\la_1, \la_3)$ and
$W_3(\la_1, \la_3)$ defining the hodograph solution (\ref{Ts1}),
\begin{equation}\label{Ts11}
 x - V_1 t = W_1  \, , \qquad   x - V_3 t = W_3   \, .
\end{equation}
Using the boundary condition  at  $x=x^+_1$ (\ref{right1}) and expression (\ref{m02})
for the characteristic velocity $V_1$  in the degenerate case when $\la_3=\la_4$,
the first  equation (\ref{Ts11}) becomes
\begin{equation}\label{hodlim1}
x- \frac{3\la_1+1}{2} t =W_1(\la_1, A^+) \, .
\end{equation}
Since according to the matching condition (\ref{right1}) one has $\la_1=\la_-$ at $x=x^+_1$ , we get, by
comparing (\ref{hodlim1}) with the rarefaction wave solution (\ref{rar1}), that
\begin{equation}\label{bcc1}
W_1(\la_1, A^+) = l \, .
\end{equation}
Next we turn to the boundary condition (\ref{left1}) and deduce from the comparison of second equation (\ref{Ts11})
with similarity solution (\ref{DSW1}) that
\begin{equation}\label{bcc2}
W_3(-1, \la_3) = 0\,.
\end{equation}
Thus, the unknown at the onset curvilinear interaction zone $PQTR$ in the $(x,t)$-plane maps to the prescribed rectangle $PQTR$
in the hodograph  $(\la_1 \la_3)$ plane (Fig.~9, right) exactly as it happens in the problem of the interaction of two simple waves in classical gas dynamics (see e.g. \cite{RY}).
We also note that, in contrast to the original free-boundary matching conditions (\ref{left1}), (\ref{right1}) for the Riemann invariants $\la_j(x,t)$,
the boundary conditions for the functions $W_{1,3}(\la_1, \la_3)$ are {\it linear} (i.e. they do not depend
on the particular solution).

%moreover, they are specified on the prescribed boundaries $\la_1=-1$ and $\la_3=A^+$ in the hodograph plane $(\la_1, \la_3)$.

To deduce  boundary conditions for the EPD equation (\ref{EPD1}) from conditions (\ref{bcc1}), (\ref{bcc2}) for the Tsarev equations (\ref{Ts2}) we use the relations (\ref{scalar})
between $W_{1,3}(\la_1, \la_3)$ and $g(\la_1, \la_3)$. Then from
 (\ref{bcc1}) we obtain a simple ODE
\begin{equation}\label{bg1}
g(\la_1, A^+) - \frac{\mathfrak{L}(\la_1,1,A^+, A^+)}{\partial_1 \mathfrak{L}(\la_1,1,A^+, A^+)} \partial_1 g(\la_1, A^+) = l \, ,
\end{equation}
which is readily integrated to give the boundary value of the function $g(\la_1, \la_3)$ at $\la_3=A^+$:
\begin{equation}\label{bg2}
g(\la_1, A^+) = C_1 \mathfrak{L}(\la_1,1,A^+, A^+) + l = \frac{C_1}{\sqrt{A^+-\la_1}}+l \, ,
\end{equation}
where $C_1$ is an arbitrary constant.

Next, from (\ref{bcc2}), (\ref{scalar}) we find
\begin{equation}\label{bg3}
g(-1, \la_3) - \frac{\mathfrak{L}(-1,1, \la_3, A^+)}{\partial_3 \mathfrak{L}(-1, \la_3)}
\partial_3 g(-1, \la_3) = 0\, ,
\end{equation}
so the solution is readily found as
\begin{equation}\label{bg4}
g(-1,\la_3)= C_2\mathfrak{L}(-1,1,\la_3, A^+)\, ,
\end{equation}
where $C_2$ is another arbitrary constant.

Conditions (\ref{bg2}) and (\ref{bg4}) represent the Goursat type characteristic boundary conditions for the EPD equation (\ref{EPD1}). Now, we have two arbitrary functions $\phi_{1,2}(\la)$
(see general solution (\ref{gs})) and two arbitrary constants $C_{1,2}$ at our disposal to satisfy boundary conditions (\ref{bcc1}) and (\ref{bcc2}).
 We first observe that, according to Section 4.4., the function $g(-1,\la_3)$ has the meaning of the modulation phase shift in the incident DSW.  Since this DSW is described by a centred simple wave modulation solution, this phase shift must be equal to zero. Thus we set $C_2=0$ so that condition (\ref{bg4}) assumes the form
\begin{equation}\label{bg5}
g(-1,\la_3)=0\,
\end{equation}
in accordance with the phase shift requirement (\ref{sh}).

Now, the easiest way to proceed is to put $\phi_2(\la)\equiv 0$ and $a_1=-1$  in (\ref{gs}) so that the solution of the EPD equation
(\ref{EPD1}) reduces to a single quadrature
\begin{equation}\label{gs1}
g=\int \limits _{-1} ^{\la_1} \frac{\phi_1(\la) d
\la}{\sqrt{(\la_3 - \la)(\la_1-\la)}} \, .
\end{equation}
Now we need to find  $\phi_1(\la)$ and $C_1$ to satisfy
two conditions (\ref{bg2}) and (\ref{bg5}).

Substitution of (\ref{gs1}) into
boundary condition (\ref{bg2}) yields
\begin{equation}\label{abel1}
\int \limits _{-1} ^{\la_1} \frac{\phi_1(\la) d
\la}{\sqrt{(A^+ - \la)(\la_1-\la)}}=\frac{C_1}{\sqrt{A^+-\la_1}}+l \, ,
\end{equation}
which is an Abel integral equation for $\phi_1(\la)$ (see e.g. \cite{abramowitz}). We recall that
\begin{equation}\nonumber
\hbox{if} \quad \int \limits^x_a\frac{\phi(\xi)}{\sqrt{x-\xi}}d\xi = f(x)\, , \quad \hbox{then} \quad \phi(x)=
\frac{1}{\pi}\frac{d}{dx}\int \limits_a^x \frac{f(\xi)}{\sqrt{x-\xi}}d \xi.
\end{equation}

Thus, the solution to (\ref{abel1}) is readily obtained in the form
\begin{equation}\label{abelsol1}
\phi_1(\la)= \frac{1}{\pi \sqrt{\la+1}} \left(C_1 \sqrt{\frac{A^+ + 1}{A^+-\la}} + l\sqrt{A^+-\la} \right)\, .
\end{equation}
Now one can see that condition (\ref{bg5}) is satisfied by (\ref{gs1}), (\ref{abelsol1}) only if $\phi_1(-1)=0$, which
implies that $C_1=-l\sqrt{A^++1}$ and so finally
\begin{equation}\label{sol1}
\begin{split}
g(\la_1, \la_3)=-\frac{l}{\pi}\int \limits_{-1}^{\la_1} \frac{\sqrt{\la+1}}{\sqrt{(A^+ - \la)(\la_3-\la)(\la_1-\la)}} d \la \\
= \frac{2l(A^++1)}{\pi \sqrt{(A^+-\la_1)(\la_3+1)}}(\Pi_1(s,z) - \K(z))\,,
\end{split}
\end{equation}
where $\Pi_1(s,z)$ is the complete elliptic integral of the third kind (see, e.g. \cite{abramowitz}) and
\begin{equation}\label{cf}
z=\frac{(A^+-\la_3)(\la_1+1)}{(A^+-\la_1)(\la_3+1)}\, , \qquad s=-\frac{\la_1+1}{A^+ - \la_1}\, .
\end{equation}
Hence, the  modulation solution describing the interaction of counter-propagating DSW and RW is given by the
formulae
\begin{equation}\label{hodsol1}
\la_2=1, \quad  \la_4=A^+, \qquad x-V_{1,3}(\la_1, 1, \la_3, A^+)t=
\left(1-\frac{\mathfrak{L}}{\partial_{1,3}\mathfrak{L}}\partial_{1,3}\right)g(\la_1, \la_3)\, ,
\end{equation}
where $g(\la_1, \la_3)$ is specified by (\ref{sol1}).
%One should note that hodograph solution (\ref{hodsol1}) specifies
%the dependencies $\la_{1,3}(x,t)$ {\it implicitly} so, strictly speaking, one should prove solvability of \ref{hodsol1}.

The interaction continues until the moment $t^*$ defined by the condition $x_2^+(t^*)=x_1^-(t^*)$ (the right edge of the RW
coincides with the trailing edge of the DSW -- the point $T$ on the diagram in Fig.~9, left). It is clear from the Riemann invariant sketch in Fig. 8 that this will take place when one has
$\la_3 = 1$ and $\la_1=A^-$ simultaneously (see also Fig.~9, right). Substituting $\la_3 = 1$ and $\la_1=A^-$  into hodograph solution (\ref{hodsol1}) we find after some algebra that
\begin{equation}\label{tstar}
t^*=\frac{2\sqrt{2}l \E(r)}{\pi(1-A^-)\sqrt{A^+ - A^-}}, \quad \hbox{where} \quad r=\frac{(A^+ - 1)(A^- +1)}{2(A^+ - A^-)}\, .
\end{equation}
The corresponding coordinate $x^*=x_2^+(t^*)=x_1^-(t^*)$ is given by
\begin{equation}\label{xstar}
x^*=\left (1+\frac{A^+ + A^-}{2} \right)t^* + P(1)\, ,
\end{equation}
 where $P(1)=g(A^-, 1)$ (to be discussed below in Section 5.4).
\subsection{After interaction, $t>t^*$}
At $t=t^*$ the DSW exits the RW region and the two waves separate.
\begin{figure}[ht]
\begin{center}
\includegraphics[width=8cm]{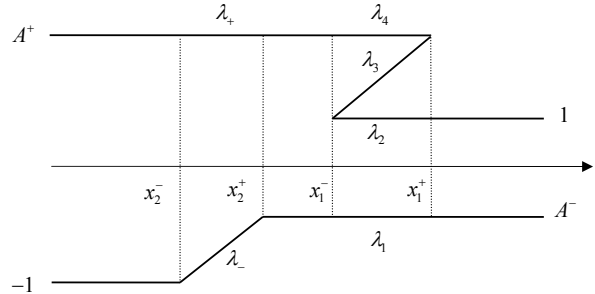}
\caption{
Schematic behaviour of the Riemann invariants after the interaction of the DSW and RW, $t>t_*$.
}
\end{center}
\end{figure}

\subsubsection{\it Refracted DSW}
The modulation solution describing the DSW after the separation is given by three constant invariants (see Fig.~10)
\begin{equation}\label{DSW2c}
\la_1=A^-\, ,\quad \la_2=1\, , \quad \la_4=A^+\, ,
\end{equation}
while for the remaining one, $\la_3$, we have from (\ref{hodsol1}) a simple-wave modulation solution (cf. (\ref{DSW1}))
\begin{equation}\label{DSW22}
\begin{split}
x=V_3(A^-,1,\la_3, A^+)t+P(\la_3) \\
\\
=\left(\tfrac12 (1+A^-+A^++\la_3) + \frac{(A^+-\la_3)(\la_3 - 1)}{(\la_3 - 1) -(\la_3 -A^-)\mu(m)} \right) t + P(\la_3)\, ,
\end{split}
\end{equation}
where
\begin{equation}\label{msim1}
m= \frac{(1-A^-)(A^+ - \la_3)}{(A^+ - 1)(\la_3 - A^-)}
\end{equation}
and the function $P(\xi)$ is found  as
\begin{equation}\label{P}
\begin{split}
P(\xi)=W_3(A^-, \xi)=
\left(1-\frac{\mathfrak{L}(A^-, 1, \xi, A^+)}{\partial_3\mathfrak{L}(A^-, 1, \xi, A^+)}
\partial_3\right)g(A^-, \xi) \\
\\
=\frac{2 l}{\pi \sqrt{(A^+ - A^-)(\xi+1)}}\left ( (A^++1)\Pi_1(p,z)+
\frac{[(A^+)^2-1](\xi - A^-)\K(z)\mu(y)-[\xi^2-1][A^+-A^-]\E(z)}{(\xi - A^-)[(\xi - 1) -(A^+-1)\mu(y)]} \right)\, ,
\end{split}
\end{equation}
where
\begin{equation}\label{sz}
 p=-\frac{A^-+1}{A^+-A^-}  \, , \qquad z= \frac{A^- +1}{A^+-A^-}\frac{A^+ - \xi}{\xi+1}\, , \qquad y=\frac{(1-A^-)(A^+ - \xi)}{(A^+ - 1)(\xi - A^-)}.
\end{equation}
Expressions (\ref{sz}) are obtained from formulae (\ref{cf}), where one sets $\la_1=A^-$, $\la_3 =\xi$,
and the modulus $m$ in (\ref{P}) is specified by (\ref{msim1}) where $\la_3$ is replaced by $\xi$.
Thus, as a result of the interaction, the DSW is no longer described by the similarity modulation solution in the form of an
expanding centred fan but rather becomes a general simple wave solution of the modulation system corresponding
to the following initial-value problem for the NLS equation (\ref{NLS}):
 \begin{equation}\label{in}
 \la_-(x,0)=A^-\, , \quad \la_+(x,0)=P^{-1}(x)\, ,
 \end{equation}
$P^{-1}(x)$ being inverse of the function $x=P(\la_+)$.
The function $P(\la_+)$ in (\ref{DSW22}) represents the DSW de-centring distribution  acquired as a result of the interaction with the RW. It is directly related to the modulation phase shift distribution $\theta_0(x,t)$ via (\ref{hodtheta}), (\ref{sh}).
It is not difficult to verify that $P(\xi) \equiv 0$ for $A^-=-1$. This is exactly what one should expect since when $A^-=-1$, there is no RW is generated and, therefore, there is no DSW refraction.

The boundaries $x_1^-$ and $x^+_1$ of the refracted DSW are found by setting in (\ref{DSW22}) $\la_3=1$ (i.e. $m=1$) and $\la_3=A^+$ (i.e. $m=0$) respectively
\begin{equation}\label{edges}
x_1^-=\left(1+ \frac{A^-+A^+}{2}\right)t+ P(1)\, , \qquad x^+_1 = \left(2A^+-\frac{(1-A^-)^2}{2(2A^+-1-A^-)}\right)t + P(A^+)\, .
\end{equation}
The  background density and the amplitude of the trailing dark soliton in the refracted DSW are (cf. (\ref{solamp1}))
\begin{equation}\label{solamp2}
n_{sr}= \frac{1}{4}(A^+ - A^-)^2\, ,  \quad a_{sr}=(A^+-1)(1-A^-)\, .
\end{equation}

The intensity $I_r$ of the refracted DSW is determined from (\ref{int}) where we set  $n_1=n_{sr}$ and $n_2=\tfrac{1}{4}(1-A^-)^2$ (the latter is defined by the initial conditions (\ref{ic2})). Thus \begin{equation}\label{ir}
I_r = \left(\frac{A^+-A^-}{1-A^-} \right)^2\, .
\end{equation}

\subsubsection{\it Refracted RW}

The solution for the refracted RW is found from the hodograph modulation solution (\ref{hodsol1}) by setting in it  $\la_4=\la_+=A^+$, $\la_3=\la_2=1$,  $\la_1=\la_-$ (see (\ref{restr1})) and using that
$V_1(\la_1, \la_3, \la_3, \la_4)=V_-(\la_1, \la_4)$ (see \ref{m1}). As a result we get
\begin{equation}\label{rar2}
\la_+=A^+\, , \qquad x=V_-(\la_-,A^+)t+G(\la_-)=\frac{3\la_-+A^+}{2}t+G(\la_-) \, ,
\end{equation}
where the function $G(\xi)$ has the form
\begin{equation}\label{G}
\begin{split}
G(\xi)= W_1(\xi, A^+)= \left(1-\frac{\mathfrak{L}(\xi, 1, 1, A^+)}{\partial_1\mathfrak{L}(\xi, 1, 1, A^+)}
\partial_1\right)g(\xi, A^+) \\
=\frac{l\sqrt{2}}{\pi\sqrt{A^+-\xi}}\left[(A^++1)(\Pi_1(n,r)-\K(r))+ 2\E(r) \right]\, ,
\end{split}
\end{equation}
where
\begin{equation}\label{nr}
r=\frac{(A^+-1)(\xi+1)}{2(A^+-\xi)}\, , \quad n=-\frac{\xi+1}{A^+-\xi}\, .
\end{equation}

Similar to the refracted DSW, the refracted RW is no longer described by a centred fan solution but rather by a general simple-wave solution of the
shallow-water system (\ref{er}), (\ref{eq20}) with the `effective' initial conditions $\la_+=1$ and $\la_-(x,0)$ given by
the function inverse to the refraction shift function $G(\la_-)$.

The boundaries of the refracted RW are given by the expressions
\begin{equation}\label{}
x_2^-=\frac{A^+-3}{2}t + G(-1) \, , \qquad x_2^+ = \frac{3A^-+A^+}{2}t + G(A^-) \, .
\end{equation}
\subsubsection{\it Vacuum points}
As already was mentioned, an important property of the DSWs in the defocusing NLS flows is the possibility of the vacuum point(s) occurrence in the solutions for the  problems not containing vacuum states in the initial data \cite{eggk95}.  This effect has no analogue in both viscous SW dynamics and in the DSW dynamics in media with negative dispersion supporting bright solitons. Across the vacuum point, the flow speed changes its sign, which implies the generation of a counterflow. As a result, the DSW with a vacuum point inside it, unlike a regular DSW, no longer represents
a single oscillatory wave of compression: the vacuum point separates the compression part propagating to the right and the {\it oscillatory rarefaction wave} propagating to the left  \cite{eggk95}. The DSW counterflow due to the vacuum point occurrence has been recently observed in  the experiments on nonlinear plane wave tunneling through  a broad penetrable repulsive potential barrier (refractive index defect) in photorefractive crystals \cite{fleischer2}.

If we fix the state $n_1=1$, $u_1=0$ in front of the DSW  (as we do for the incident wave), then, by increasing the density jump $n_2$ across the DSW we will be able to increase the DSW relative intensity only up to the value $I=4$ at which the vacuum point
occurs at the DSW trailing edge \cite{gk87}. If  $n_2$ increases further, beyond the vacuum point threshold, the relative intensity of the compression part of the DSW decreases  and, asymptotically as $n_2/n_1 \to \infty$, vanishes so that the DSW completely transforms into the classical (smooth) left-propagating rarefaction wave  \cite{eggk95}. This limit can alternatively be achieved by keeping the upstream state $n_2$ fixed and  letting $n_1 \to 0$: then we arrive at the  well-known solution of the classical shallow-water dam-break problem (see e.g. \cite{whitham74}).

Setting $A^-=-1$ we recover the already mentioned  criterion $A^+ \ge 3$ for the vacuum point occurrence in the incident DSW.
If $a_{sr} = n_{sr}$, which by (\ref{solamp2}), yields the relation $A^+=2-A^-$, then the condition for the vacuum point appearance in the refracted DSW assumes the form
\begin{equation}\label{apmcrit}
A^+ \ge 2-A^- \, .
\end{equation}
The regions of the $A^-, A^+$ plane corresponding to different (with respect to the vacuum point appearance) flow configurations arising in the initial-value problem (\ref{NLS}), (\ref{ic1}) are presented in a diagram shown in Fig.~11. A particular flow evolution corresponding to Region II is shown in Fig.~12.
\begin{figure}[htp]
\begin{center}
\includegraphics[width=5cm]{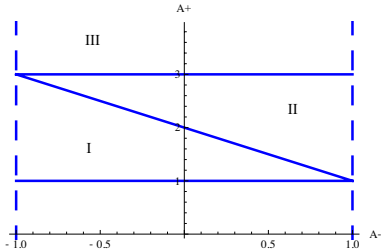}
\caption{Regions in the plane of initial parameters $(A^-, A^+)$ --- the classification with respect to the vacuum point occurrence. (I):  No vacuum points; (II): No vacuum points in the incident DSW, a vacuum point in the refracted DSW;
(III): Vacuum points in both incident and refracted DSWs.
}
\end{center}
\end{figure}
We stress that, although the vacuum point appearance modifies the oscillatory DSW profile (the lower DSW density envelope becomes nonmonotonous and the velocity profile acquires a singularity at the vacuum point --- see \cite{eggk95}, \cite{ha06}), all the dependencies of  the DSW edge speeds, density jumps and trailing soliton amplitudes on the initial data $A^+, A^-$ remain unchanged.
\begin{figure}[htp]
\begin{center}
\includegraphics[width=16cm, clip]{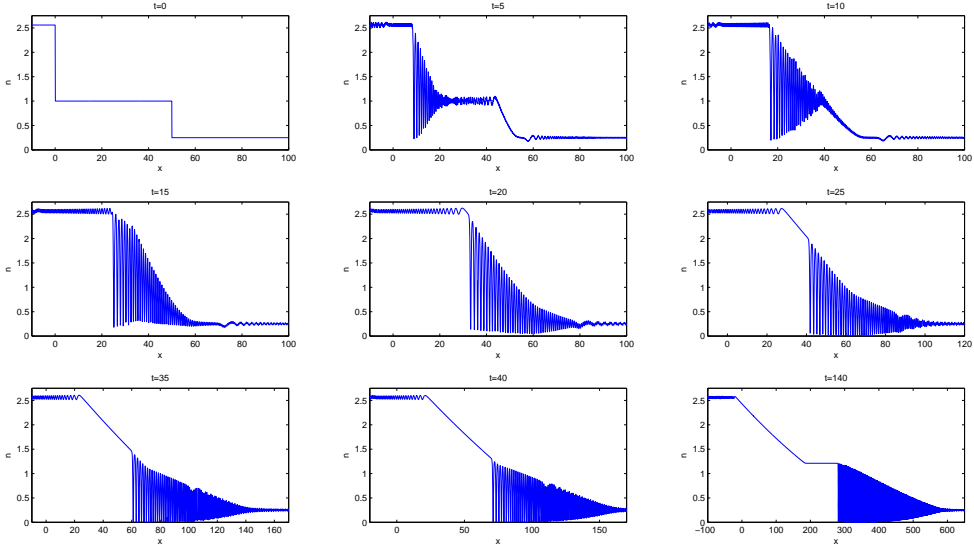} \\
\caption{Evolution of the profile (\ref{ic1}) with $A^-=0$, $A^+=2.2$, $l=50$ (Region II in Fig.~11) leading to the occurrence of a vacuum point in the refracted DSW. }
\label{fig12}
\end{center}
\end{figure}

\subsection{Key parameters of DSW refraction}

It is convenient to characterise the DSW refraction  by three key parameters: the  amplification coefficient $\nu$ which culd be defined as the ratio of the relative intensities (\ref{int}) of the refracted and the incident DSWs, the acceleration coefficient $\sigma$ which we define as the difference  between the values of the DSW trailing dark soliton speeds $s^-$ after and before the interaction, and the refraction  shift $d$ which is naturally defined as the phase shift of the DSW trailing soliton  due to the DSW interaction with the RW (see Figs.~4,9).

For the first two parameters we readily have:
\begin{equation}\label{nu}
\nu= \frac{I_r }{I_0}=  \left( \frac{2(A^+-A^-)}{(1-A^-)(1+A^+)} \right)^2\,
\end{equation}
--- see (\ref{ir}), (\ref{i0}), and
\begin{equation}\label{sigma}
\sigma =s_r-s_0 = \left.\frac{dx_1^-}{dt}\right|_{t>t^*} - \left.\frac{dx_1^-}{dt}\right|_{t<t_0} =\frac{1+A^-}{2} >0\, .
\end{equation}
-- see (\ref{edges}), (\ref{bound1}).

Here the subscripts `$0$' and `$r$' refer to the incident and refracted waves respectively. Note that the determination of $\nu$ and $\sigma$ actually does not require knowledge of the full solution: both quantities are determined by the the transfer of the Riemann invariants through the DSW region.
Interestingly, the acceleration  coefficient $\sigma$ does not depend on the  DSW strength  before the interaction ($ \sim A^+$)  and is completely determined by the initial jump $A^-$ of the Riemann invariant $\la_-$ across the RW.
It also follows from (\ref{sigma}) that, since $A^->-1$, one has $\sigma >0$  i.e. the DSW is always accelerated as a result of the head-on collision with the RW (indeed, $\sigma >0$ implies acceleration of the trailing edge of the DSW and, therefore, acceleration of the DSW as a whole).
The SW acceleration in the head-on collision with RW  is also always the case in classical gas dynamics (see, e.g. \cite{moses}) as the SW meets the gas of decreasing density.

Unlike the acceleration coefficient $\sigma$, the amplification coefficient $\nu$ can have both signs depending on the specific values of $A^+$ and $A^-$ chosen, the boundary between the regions of the DSW (relative) strengthening and attenuation being given by equation $A^+=(1-A^-)/(1+A^-)$. We also note that, while the amplification coefficient $\nu$ is formally defined for the full range of values of
$A^+$ and $A^-$, its original significance is retained only for the DSWs not containing vacuum points (see the discussion in the previous Section).

The function (see (\ref{P}), (\ref{edges}))
\begin{equation}\label{refshift}
d(A^+, A^-)=P(1) =\frac{\sqrt{2} l}{\pi}\frac{ (A^++1)}{ \sqrt{(A^+ - A^-)}}\left ( \Pi_1(p,z^*)
- \K(z^*)  \right)\, ,
\end{equation}
where (see  (\ref{sz}))
\begin{equation}\label{pz}
z^*= \frac{A^- +1}{A^+-A^-}\frac{A^+ - 1}{2}\, , \qquad p=-\frac{A^-+1}{A^+-A^-}\, ,
\end{equation}
describes the {\it refraction shift} (see Fig.~4) of the trailing dark soliton in the DSW as a function of the initial parameters $A^+$, $A^-$.  As a matter of fact, the determination of the refraction phase shift {\it does require} the knowledge of the full modulation solution in the interaction region. One can observe by comparing (\ref{refshift}) with solution $g(\la_1, \la_3)$ (\ref{sol1}), (\ref{cf}) of the
EPD equation for the DSW-RW interaction region, that
\begin{equation}\label{}
d(A^+, A^-)=g(A^-, 1)\, ,
\end{equation}
which corresponds to the value of $g$ at the moment $t=t^*$ (see (\ref{tstar})), when the DSW exits the interaction region
--- see Figs.~8, 9. This is, of course,  expected from the general modulation phase shift consideration described in Section 4.4.

The dependencies of the refraction phase shift $d$  on $A^-$ and $A^+$ given by (\ref{refshift}), along with  direct numerical simulations data for the refraction shift, are presented in Fig.~13.  One can see that the dependence of the refraction shift on the density jump across the RW (roughly proportional to the value of  $A^-$) is much stronger than on the incident DSW strength (proportional to $A^+$). Along with the curves for the exact analytical solution for $d$, we also present the values of $d$ extracted from the numerical simulations. One can see that, despite the fact that the accuracy of the modulation solution in the interaction region is not expected to be very high for the moderate spacing $l$ between the initial jumps for the dispersionless Riemann invariants $\la_{\pm}$, the agreement between the asymptotic solution and direct numerics is quite good.
The plots and comparisons with numerics for $\sigma$ and $\nu$ will be presented in the next section as particular cases in the study of the DSW-RW interaction  in  the framework of a generalised, non-integrable version of the NLS equation.
\begin{figure}[h]
\begin{center}
\includegraphics[width=5cm]{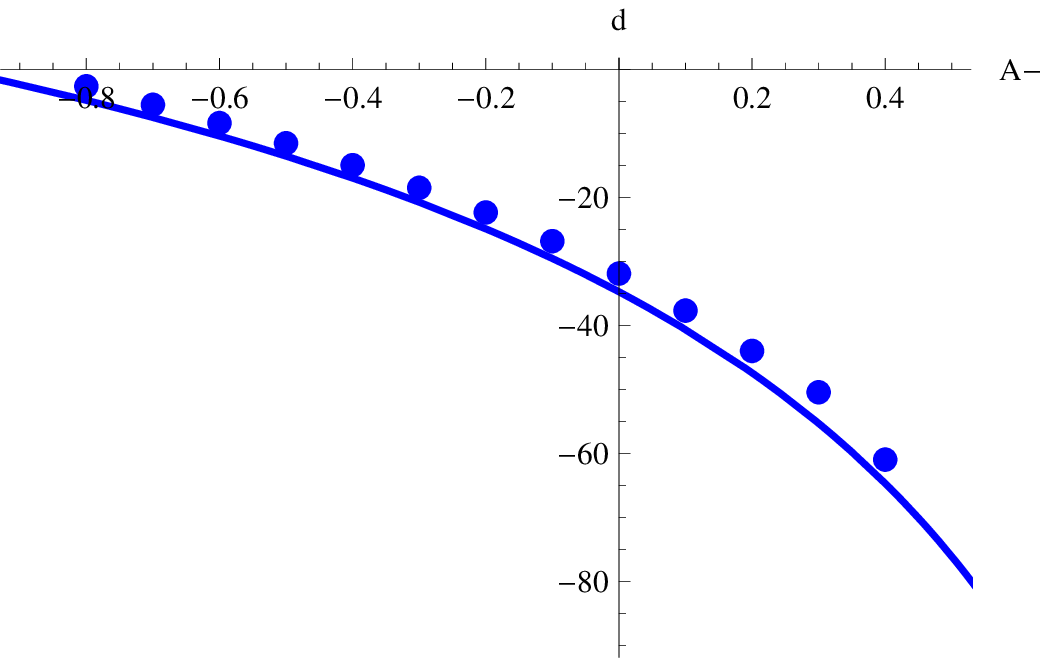} \qquad \includegraphics[width=5cm]{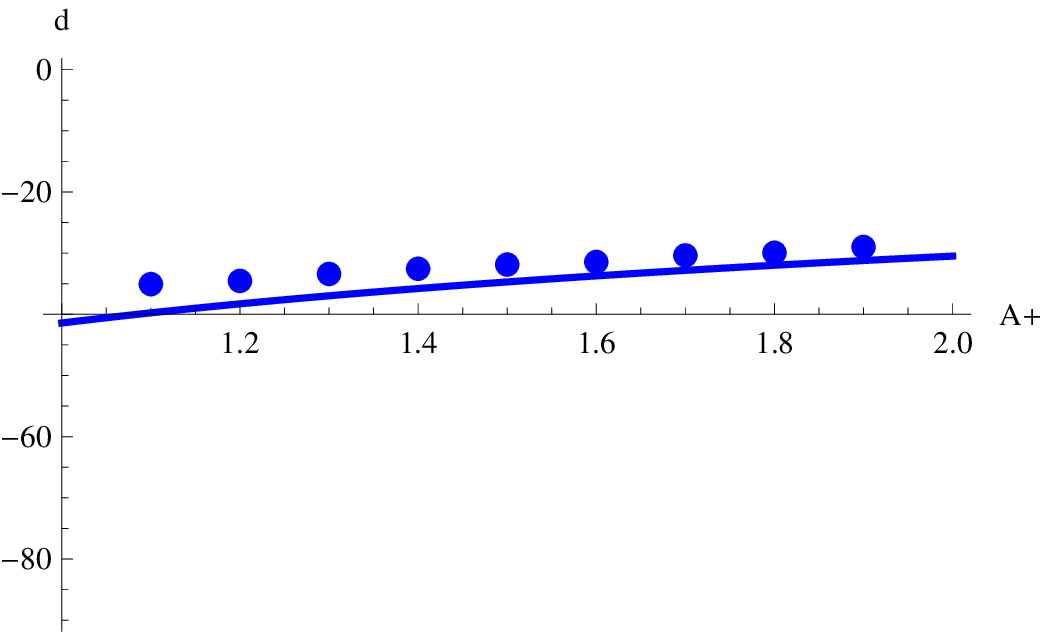}
\caption{Typical behaviour of the DSW refraction phase shift $d$. Left: dependence $d(A^-)$ for fixed $A^+=1.5$, right: dependence $d(A^+)$ for fixed $A^-=0$. Solid line: formula (\ref{refshift}), circles: direct numerical simulations data.}
\end{center}
\end{figure}

One can trace certain analogy between the considered DSW-RW interaction and the two-soliton collisions in integrable systems:  both  interactions are elastic in the sense that they both can be interpreted in terms of the ``exchange'' of spectral parameters by the interacting waves  so that the global spectrum in the associated linear scattering problem remains unchanged. In the DSW-RW interaction the role of  isospectrality is played by the transfer of the constant values of appropriate  Riemann invariants of the modulation system through the varying DSW and RW regions so that one can predict the jumps of density and velocity across the refracted DSW and RW without constructing the full modulation solution. At the same time, the DSW and RW do not simply pass through each other and ``exchange'' the constant Riemann invariants: there are additional phase shifts for both interacting waves, similar to the classical soliton phase-shifts. The determination of these phase shifts requires knowledge of the full modulation solution.

\section{Refraction of dispersive shock waves in optical media with saturable nonlinearity}

\subsection{Formulation of the problem}

We now consider the NLS equation with saturable nonlinearity (hereafter called the sNLS equation)
\begin{equation}\label{snls}
    i \psi_t+\frac1{2}\psi_{xx}-
    \frac{|\psi|^2}{1+\gamma|\psi|^2}\psi=0,
\end{equation}
where $\gamma>0$ is the saturation parameter. This equation describes, in a certain approximation,  the one-dimensional propagation of a plane stationary light beam through
a photo-refractive crystal  (see e.g.  \cite{gatz91}, \cite{christ95}). One should note that in the nonlinear optics context the role of the time
variable $t$ is played by the spatial coordinate $z$ along the beam propagation direction while $x$ is the transversal coordinate.
If the saturation
effect is negligibly small ($\gamma|\psi|^2\ll1$), then the sNLS equation (\ref{snls})
reduces to the cubic NLS equation (\ref{3-1}).
The Madelung transformation (\ref{3-2}) with $\epsilon=1$ maps equation (\ref{snls}) to the dispersive hydrodynamics system (cf. (\ref{NLS})),
\begin{equation}\label{sNLS}
\begin{split}
    n_t+(nu)_x=0,\\
 u_t+uu_x+\left(\frac{n}{1+\gamma n}\right)_x+\left(\frac{n_x^2}{8n^2}
   -\frac{n_{xx}}{4n}\right)_x=0\, .
   \end{split}
\end{equation}
Here $n$ has the meaning of the light beam intensity and $u$ is the local value of the wave vector component transversal to the beam propagation direction. A detailed study of the periodic solutions to (\ref{sNLS}) can be found in \cite{egkk07}.
In particular, the linear dispersion relation  for the waves of infinitesimally small amplitude propagating against the constant background flow with
$u=u_0$, $n=n_0$ has the form
\begin{equation}\label{dr1}
\omega = \omega_0(n_0, u_0,k)=ku_0 \pm k
\sqrt{\frac{n_0}{(1+\gamma n_0)^2} +\frac{k^2}{4}} \, ,
\end{equation}
where $\omega$ is the wave frequency and $k$ is the wavenumber.

In the dispersionless limit, system (\ref{sNLS}) can be cast in the diagonal form (\ref{er}) with the Riemann invariants $\la_\pm$ and characteristic velocities $V_{\pm}$
expressed in terms of the hydrodynamic variables $n$ and $u$ as
\begin{equation}\label{srim}
    \la_{\pm}=\frac{u}{2} \pm \frac{1}{\sqrt{\gamma}}\arctan \sqrt{\gamma n}, \quad V_{\pm}=u \pm \frac{\sqrt{n}}{1+\gamma n}\, .
\end{equation}
When $\gamma \to 0$ expressions (\ref{srim}) go over to the shallow-water relationships (\ref{eq20}), (\ref{V}) (note
the different normalization for the dispersionless Riemann invariants compared to that used in \cite{egkk07}).

Similar to (\ref{ic1}),  we specify the initial conditions for (\ref{sNLS}) in terms of two steps for the hydrodynamic Riemann invariants $\la_{\pm}$
\begin{equation}\label{ics1}
\la_+(x,0) =\left\{
\begin{array}{ll}
A^+  &\quad \hbox{for} \quad x <0,\\
\frac{1}{\sqrt{\gamma}}\arctan \sqrt{\gamma} & \quad \hbox{for} \quad x>0;
\end{array}
\right.
\qquad  \la_-(x,0)=\left\{
\begin{array}{ll}
 - \frac{1}{\sqrt{\gamma}}\arctan \sqrt{\gamma} & \quad \hbox{for} \quad x<l, \\
 A^-  &\quad \hbox{for} \quad x >l,
\end{array}
\right.
\end{equation}
where $A^+> \frac{1}{\sqrt{\gamma}}\arctan \sqrt{\gamma}$, and $-\frac{1}{\sqrt{\gamma}}\arctan \sqrt{\gamma}<A^-<\frac{1}{\sqrt{\gamma}}\arctan \sqrt{\gamma}$. The special values of
$\la_+$ for $x>0$ and $\la_-$ for $x<l$ are chosen such that initially the DSW and RW will propagate into an undisturbed ``gas''
(indeed, one can readily see that $n=1$, $u=0$ in the middle region $0<x<l$ (cf. (\ref{ic2})).

Our numerical  simulations of the evolution (\ref{sNLS}), (\ref{ics1}) showed that, for a broad range of initial data parameters $A^\pm$, the {\it qualitative} DSW refraction scenario is the same as in the cubic NLS case studied in previous sections. The {\it quantitative} characteristics of the DSW refraction, however, now depend not only on the initial conditions but also on the saturation parameter $\gamma$ entering the sNLS equation. This dependence  was shown in \cite{egkk07} to be quite strong for isolated photorefractive DSWs.  We mention that knowledge of the effects of the photorefractive saturation on the parameters of a DSW  is especially important in the context of an all-optical modelling of  BEC  dynamics (see \cite{fleischer07}).  Thus the DSW-RW interaction for the sNLS equation deserves a special study.

Since the sNLS equation (\ref{sNLS}) is not integrable by the IST, the Riemann invariants are not available for the associated Whitham system and the modulation solution cannot be constructed by the methods used in previous Sections. An analytic description of the DSW refraction requires now a different technique. We shall take advantage of the theory of DSWs in photorefractive media developed in \cite{egkk07} and based on the `dispersive shock fitting'
method introduced in \cite{el05}. As already was mentioned, our specific interest here is to quantify the effect of the nonlinear saturation on the DSW refraction, and, in particular, on the parameters  $\sigma$ and $\nu$ introduced  above in the cubic NLS context (see
(\ref{sigma}), (\ref{nu})).
\subsection{DSW transition relations}
The key ingredients of the dispersive shock fitting method of \cite{el05} in  application to the  sNLS equation (\ref{sNLS}) can be formulated as follows (see  \cite{egkk07} for the details pertinent to the present study). Let the {\it right-propagating} DSW be confined to a
finite region of space $x^-<x<x^+$ and connect two constant hydrodynamic states $(n_1, u_1)$ at $x<x^-$ and $(n_2, u_2)$ at $x>x^+$; $n_1>n_2$. Such a DSW is called a simple DSW. At the trailing edge $x^-$ the simple DSW assumes the form of a dark soliton moving with constant velocity $s^-$ and at the leading edge $x^+$ it degenerates into a vanishing amplitude linear wavepacket moving with constant group velocity $s^+$, $s^+ > s^-$. The lines $x^{\pm}=s^{\pm}t$ represent free boundaries where  the continuous matching of the mean flow $(\bar n, \bar u)$ in the DSW region with the external constant states $(n_1,u_1)$ and $(n_2,u_2)$ occurs (in some cases it is more advantageous  to formulate the matching conditions in terms of the mean density $\bar n$ and the mean momentum $\overline{nu}$ --- see e.g. \cite{ha06}).

The simple DSW transition between the hydrodynamic states $(n_1,u_1)$ and $(n_2,u_2)$ is described by the following relationships:
\begin{itemize}
\item The value of the Riemann invariant $\la_-$ is conserved across the  DSW ,
\begin{equation}\label{tran}
\left.\la_- \right |_{x=x^-} = \left.\la_- \right |_{x=x^+}\, ,
\end{equation}
i.e.
\begin{equation}\label{c1}
\frac{u_1}{2} - \frac{1}{\sqrt{\gamma}}\arctan \sqrt{\gamma n_1} = \frac{u_2}{2} - \frac{1}{\sqrt{\gamma}}\arctan \sqrt{\gamma n_2}  \equiv \la_-^0 \, .
\end{equation}

\item The DSW edge speeds $s^{\pm}$ are defined by the kinematic conditions (cf. conditions (\ref{mult}) the cubic NLS case)
\begin{equation}\label{spm}
s^+=\left.\frac{\partial \Omega}{\partial k}\right|_{\bar n =n_2, \ k=k^+} \, ;  \quad \qquad \left.s^-=\frac{\widetilde \Omega}{\kappa}\right|_{\bar n =n_1, \ \kappa=\kappa^-}\,.
\end{equation}
The quantities $k^+$ (the leading edge wavenumber) and $\kappa^-$ (the trailing edge ``soliton wavenumber" -- the trailing soliton inverse half-width) in (\ref{spm}) represent the boundary values, $k^+=k(n_2)$ and $\kappa^-=\kappa(n_1)$, of  two functions $k(\bar n)$ and $\kappa(\bar n)$ satisfying the following ordinary differential equations (ODEs):
\begin{equation}\label{ODEk}
\frac{dk}{d \bar n}=\frac{\prt\Omega/\prt \bar n}{v_+(\bar n)-\prt\Omega/\prt
    k}, \qquad k(n_1)=0\,;
\end{equation}
\begin{equation}\label{ODEkap}
    \frac{d\kappa}{d \bar n}=\frac{\prt\widetilde{\Omega}/\prt\bar n}
    {v_+(\bar n)-\prt\widetilde{\Omega}/\prt \kappa}\, , \qquad \kappa(n_2)=0\, .
\end{equation}
Here
\begin{equation}\label{vplus}
v_+(\bar n) = V_+( \bar n, \bar u(\bar n)) = \bar u (\bar n)+ \frac{\sqrt{\bar
n}}{1+\gamma \bar n}\,\, ,
\end{equation}
\begin{equation}\label{Omega}
    \Omega(\bar n,k)=\omega_0(k, \bar u (\bar n), \bar n)=k\left[\bar u(\bar n)
   +\sqrt{\frac{\bar n}{(1+\gamma \bar n)^2}+
    \frac{k^2}4}\right]\, , \qquad \widetilde\Omega(\bar n, \kappa)=-i\Omega(\bar n, i\kappa);
\end{equation}
and
\begin{equation}\label{un}
\bar u(\bar n)= 2 \left(\la_-^0 +\frac{1}{\sqrt{\gamma}}\arctan \sqrt{\gamma \bar n} \right)\, .
\end{equation}

\item ``Entropy'' inequalities must hold ensuring that the hydrodynamic characteristics transfer data {\it into} the DSW region:
\begin{equation}\label{entropy}
    V^1_- <s^- < V^1_+, \quad V^2_+<s^+,\quad s^+ > s^-.
\end{equation}
Here $ V^1_\pm \equiv V_{\pm}(n_1, u_1)$, $V^2_+ \equiv V_+(n_2, u_2)$ -- see (\ref{srim}) for the definitions of $V_{\pm}(n,u)$.
We note that inequalities (\ref{entropy}) represent the dispersive-hydrodynamic analogs of classical Lax's entropy conditions \cite{lax}.
\end{itemize}
 Relationships (\ref{tran}) -- (\ref{entropy}) enable one to `fit' the DSW into the solution of the dispersionless limit equations  without the knowledge of the detailed solution of the full dispersive system within the DSW region (much as in classical gas dynamics SW is fitted into the solution of the inviscid equations by means of the Rankine-Hugoniot conditions subject to Lax's entropy condition).

 Using the speed-amplitude relationship for the photorefractive dark solitons obtained in \cite{egkk07} one can find the amplitude $a_s$ of the DSW trailing soliton. Setting the value $s^-$ (\ref{spm}) of the DSW trailing edge for the soliton velocity $c$ in  formula (39) of \cite{egkk07} we obtain
 \begin{equation}\label{speedamp}
(s^--u_1)^2=\frac{2(n_1-a_s)}{\gamma a_s}\left[\frac{1}{\gamma a_s}\ln
\frac{1+\gamma n_1}{1+\gamma (n_1-a_s)}- \frac{1}{1+\gamma n_1}
\right]
\end{equation}
(note: $u_1(n_1)$ is given by the simple DSW transition condition (\ref{c1})).
\subsection{DSW refraction}
Our concern in this section will be with the calculations of two DSW refraction parameters: the DSW amplification  and acceleration coefficients, defined earlier in (\ref{sigma}) and (\ref{nu}) as
\begin{equation}\label{sigma1}
\nu= {I_r}/{I_0}\,  \qquad \hbox{and} \qquad \,  \sigma=  s^-_r - s^-_0\,
\end{equation}
respectively. We note that analytical determination of the refraction phase shift $d$ is, unfortunately, not feasible now as it requires knowledge of the full modulation solution, which is not available  for the sNLS equation due to its nonintegrability so we shall present only
numerical results for $d$.

\subsubsection{Before interaction, $t < t_0$}
The previous analysis of \cite{egkk07} suggests  that the decay of two spaced initial discontinuities (\ref{ics1}) for the hydrodynamic Riemann invariants $\la_{\pm}$ would result,  similar to the cubic NLS case, in a combination of a right-propagating simple DSW centred at $x=0$ and a left-propagating simple RW centred at $x=l$.
Indeed, the simple DSW transition condition (\ref{c1}) is satisfied by the initial step at $x=0$, which implies a single DSW  resolution of this step (provided the ``entropy conditions'' (\ref{entropy}) are satisfied -- see \cite{egkk07} for the justification); similarly, the jump at $x=l$ with constant Riemann invariant $\la_+$ across it asymptotically produces a single left-propagating RW (see Fig.~14).  Indeed, our  numerical simulations of the sNLS equation (\ref{sNLS}) for a range of the
saturation parameter $\gamma$ values confirm this scenario producing the plots qualitatively equivalent to that presented in
Fig.~3.

Now, following \cite{egkk07}, we derive the key parameters of the simple photorefractive DSW in the  form convenient for the further application to the refraction problem.
\begin{figure}[h]
\begin{center}
\includegraphics[width=10cm]{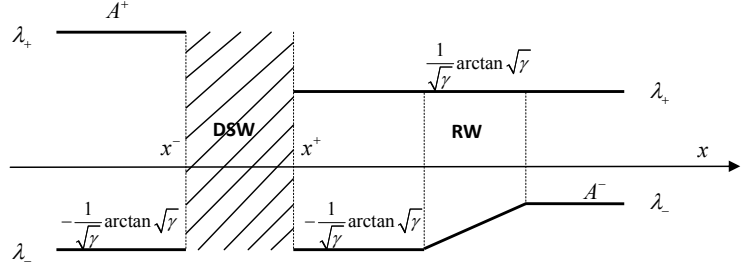}
\caption{Distribution of the classical (dispersionless limit) Riemann invariants before the DSW-RW interaction}
\end{center}\label{fig14}
\end{figure}

To take advantage of  formulae (\ref{spm}) -- (\ref{un}) for the speeds of the DSW edges we first need to find the constant states $(n_{1},u_1)$ at $x<x^-$ and $(n_2, u_{2})$ at $x>x^+$ defining the hydrodynamic jumps across the DSW. These are readily found from the  the initial conditions (\ref{ics1}) and the relationship (\ref{c1}) for the transfer of the Riemann invariant $\la_-$ across the simple DSW.
According to the initial conditions (\ref{ics1}) the simple DSW must connect two hydrodynamic states with
the same $\la_-=- \frac{1}{\sqrt{\gamma}}\arctan \sqrt{\gamma}$ while $\la_+=A^+$ for $x<x^-$ and $\la_+=\frac{1}{\sqrt{\gamma}}\arctan \sqrt{\gamma}$ for $x>x^+$ (see Fig.~14). Then, using (\ref{c1}) and expressions (\ref{srim}) relating the Riemann invariants and the hydrodynamic variables $n,u$ we find
\begin{equation}\label{n2u2}
n_2=1, \quad u_2=0\, , \quad n_1= \frac{1}{\gamma} \tan^2 \left(\frac{A^+\sqrt{\gamma}+\arctan \sqrt{\gamma}}{2} \right)\, , \quad u_1=A^+-\frac{1}{\sqrt{\gamma}}\arctan \sqrt{\gamma} \, .
\end{equation}
Thus, the $I_{0}$ of the incident DSW defined by (\ref{int}) is simply
\begin{equation}\label{io}
I_0=\frac{1}{\gamma} \tan^2 \left(\frac{A^+\sqrt{\gamma}+\arctan \sqrt{\gamma}}{2} \right)\, .
\end{equation}

Next, from (\ref{c1}) we have $\la_-^0= - \frac{1}{\sqrt{\gamma}}\arctan \sqrt{\gamma}$, which by (\ref{un}) yields
$
\bar u (\bar n)=\frac{2}{\sqrt{\gamma}}(\arctan{\sqrt{\gamma
\bar n}}- \arctan{\sqrt{\gamma}})
$
and so completely defines, via (\ref{vplus}), (\ref{Omega}), ODEs (\ref{ODEk}), (\ref{ODEkap}).

As was shown in \cite{egkk07}, it is convenient to introduce a new variable $\widetilde{\alpha}$ instead of $\kappa$ using the substitution
\begin{equation}\label{alpha}
    \widetilde{\alpha}=\sqrt{1-\frac{\kappa^2(1+\gamma \bar n)^2}{4 \bar n}}\, ,
\end{equation}
which reduces ODE (\ref{ODEkap}) to the form
\begin{equation}\label{ODEalpha}
    \frac{d\widetilde{\alpha}}{d \bar n}=-\frac{(1+\widetilde{\alpha})
    [1+3\gamma \bar n+2\widetilde{\alpha}(1-\gamma \bar n)]}{2\bar n(1+\gamma \bar n)
    (1+2\widetilde{\alpha})}\, , \qquad \widetilde{\alpha}(1)=1.
\end{equation}
The form (\ref{ODEalpha}) has an advantage of being a separable ODE when $\gamma=0$, which makes it especially useful for the asymptotic analysis for small $\gamma$ .
Once the function $\widetilde{\alpha}(\bar n)$ is found, the velocity
of the trailing soliton is determined by Eqs.~(\ref{spm}),
(\ref{Omega}) as
\begin{equation}\label{smin1}
    s^-_0=\frac2{\sqrt{\gamma}}(\arctan\sqrt{\gamma n_1}-\arctan\sqrt{\gamma})
    +\frac{\sqrt{n_1}}{1+\gamma n_1}\widetilde{\alpha}(n_1)\, ,
\end{equation}
where $n_1$ is given by Eq. (\ref{n2u2}).

The amplitude of the trailing soliton is given by speed-amplitude relationship (\ref{speedamp}). Using (\ref{speedamp}), (\ref{smin1}) and the relationship $u_1=\tfrac2{\sqrt{\gamma}}(\arctan\sqrt{\gamma n_1}-\arctan\sqrt{\gamma})$ following from (\ref{c1}) one can derive the condition of the vacuum point occurrence at the DSW trailing edge (see \cite{egkk07}):
\begin{equation}\label{vc}
\widetilde{\alpha}(n_1)=0
\end{equation}
Condition (\ref{vc}) yields, for a given value of the saturation parameter $\gamma$, the  value of the initial density jump  $n_1$ (and, therefore, of the parameter $A^+$ --- see (\ref{n2u2}))
corresponding to the vacuum point appearance at the DSW trailing edge. Say, for $\gamma=0.2$ this value of $A^+$ is about $2.18$ (cf. the critical value $A^+=3$ for $\gamma=0$)

In conclusion of this Section we present an asymptotic expansion of $s^-_0$ for small $\gamma$.
First, to leading order we have from (\ref{ODEalpha}) a separable ODE
\begin{equation}\label{ODE1gamma0}
 \gamma=0: \qquad   \frac{d\widetilde{\alpha}}{d \bar n}=-\frac{1+\widetilde{\alpha}}{2 \bar n}\, , \quad \widetilde{\alpha}(1)=1\, ,
\end{equation}
which is readily integrated to give
\begin{equation}\label{14-5}
    \widetilde{\alpha}(\bar n)=\frac2{\sqrt{\bar n}}-1 \equiv \widetilde{\alpha}_0(\bar n).
\end{equation}
%Then passing to the limit as $\gamma \to 0$ in (\ref{smin1}) and denoting the limit as $s^-_0$ we recover the cubic NLS result %(\ref{bound1})
%\begin{equation}\label{smin0}
%    s^-_0=\frac{1+A^+}{2}.
%\end{equation}
We now introduce
\begin{equation}\label{a1}
\widetilde{\alpha}=\widetilde{\alpha}_0+ \widetilde{\alpha}_1 .
\end{equation}
Substituting (\ref{a1}) into (\ref{ODEalpha})  and assuming $\widetilde{\alpha}_1 \sim \gamma$ for $\gamma \ll 1$ we obtain to first order
\begin{equation}\label{102}
    \frac{d\widetilde{\alpha}_1}{d\bar n}=-\frac{\widetilde{\alpha}_1}{2 \bar n}
    +\frac{4-3\sqrt{\bar n}}{4 - \sqrt{\bar n}}\frac{2 \gamma}{\sqrt{\bar n}}.
    \qquad \widetilde{\alpha}_1(1)=0,
\end{equation}
Eq. (\ref{102}) is  readily integrated to give
\begin{equation}\label{alpha1}
\widetilde{\alpha}_1(\bar n)=\frac{2\gamma}{\sqrt{\bar n}}\left(3(\bar n-1)+16(\sqrt{\bar n}-1)
+64\left[\ln\frac{4-\sqrt{\bar n}}3
\right]\right) \, .
\end{equation}
Now, substituting (\ref{a1}), (\ref{alpha1}) into (\ref{smin1}) and using expansion of $n_1$ (\ref{n2u2})
for small $\gamma$ we obtain to first order
\begin{equation}\label{smin3}
   s^-_0=\frac{A^++1}{2}+\gamma \left(\frac{1}{12}[(A^+)^3+15(A^+)^2+219A^+-245]+128\ln{\frac{7-A^+}{6}}\right) + O(\gamma^2)\, .
\end{equation}

As one can see, expression (\ref{smin3}) agrees to leading order with the cubic NLS result (\ref{bound1}) for the trailing edge speed. We also notice that our perturbation approach formally breaks down for $A^+ \ge 7$ because of the logarithmic divergence in Eq.~(\ref{smin3}) as $A^+ \uparrow 7$ (we note that
such values of $A^+$ correspond to very large density jumps ($n_1/n_2>10$) across the DSW --- see \cite{egkk07}).

Formulae (\ref{n2u2}), (\ref{smin1}) defining all the key parameters of the simple photorefractive DSW have been compared in \cite{egkk07} with
direct numerical simulations data for a wide range of values of the saturation parameter $\gamma$ and a very good agreement was
found.

\subsubsection{After interaction, $t > t^*$}
 Relations (\ref{tran}) -- (\ref{entropy}) describe a simple DSW transition between two {\it constant} states so they are not applicable to the varying transition in DSW-RW interaction region.  However, one should still be able to use these relations for the determination of the key parameters of the refracted DSW when the interaction is over, provided no new {\it hydrodynamic} waves
 (DSWs or RWs) are generated and the output pattern consists only of a pair of the refracted  DSW and RW separated by a constant flow.   If this is the case,
one can say that the DSW-RW interaction is hydrodynamically (or semi-classically) ``clean'' (elastic). We stress that some  zero-mean radiation due to non-integrability of the sNLS equation may be present but the latter does not affect the hydrodynamic transition conditions across the refracted DSW and RW.

The notion of a hydrodynamically clean DSW-RW interaction can be elucidated by revisiting the defocusing cubic NLS equation case considered in the previous sections.  The ``elasticity'' of the DSW refraction for this case  can be deduced from the following properties of the initial-value problem for the defocusing NLS equation with a piecewise-constant initial  datum (\ref{ic2}): (i) the asymptotic ($\epsilon \ll 1$) solution  at any moment can only contain genus-zero (RW) or genus-one (single-phase DSW) regions, so in the semi-classical limit it can be globally described by the single-phase averaged NLS-Whitham equations -- see \cite{bk06}; (ii) the defocusing NLS-Whitham system is hyperbolic; (iii) the unique combination of the output (refracted) waves is determined by the transfer of the appropriate dispersionless limit Riemann invariants across the genus-one (DSW) and  genus-zero (RW) regions. As a result, the ``clean'' DSW-RW interaction diagram on the $x$-$t$ plane has the form shown in Fig.~9 (left).

We note that rigorous justifications of properties (i), (ii) (see \cite{bk06})  are based on the presence of the integrable structure  whereas  property (iii) can be deduced using the classical method of characteristics for hyperbolic hydrodynamic type systems and does not require the availability of the Riemann invariants for the Whitham system and, hence,  does not rely on integrability of the original equation (see \cite{el05}).

Since the sNLS equation is not integrable we can only assume properties (i) and (ii) and  then apply the transition relation (\ref{c1}) (property (iii)) to the refracted DSW to determine the values  $n=n_1$ and $u=u_1$ in the `plateau' region between the refracted DSW and RW. Then our assumptions (i) and (ii) in the context of the sNLS equation could be indirectly justified {\it a-posteriori} by the comparison of the analytically obtained $n_1$ and $u_1$  with the corresponding density and velocity in the direct numerical solution of the IVP (\ref{sNLS}), (\ref{ics1}).

Since the refracted DSW propagates to the right, into the region with $\la_-=A^-$ (see the initial conditions (\ref{ics1}) at $x \to + \infty$) one must have, by (\ref{tran}), the same $\la_-=A^-$ across it, in the constant `plateau' region. Next, the refracted RW  propagates to the left, into the region with  $\la_+=A^+$ (again, see initial conditions (\ref{ics1}) at $x \to - \infty$) and, therefore, $\la_+=A^+$ everywhere through this wave and in the `plateau' region.
 From the initial condition (\ref{c1}), the value of $\la_-$ to the left of the RW is $\la_-=-\frac{1}{\sqrt{\gamma}}\arctan \sqrt{\gamma}$ and the value of $\la_+$ to the right of the DSW is $\la_+=\frac{1}{\sqrt{\gamma}}\arctan \sqrt{\gamma}$. Thus, we arrive at the Riemann invariant diagram schematically shown in Fig.~15 (cf. diagram
 in Fig.~10 for the cubic NLS case).

\begin{figure}[h]
\begin{center}
\includegraphics[width=10cm]{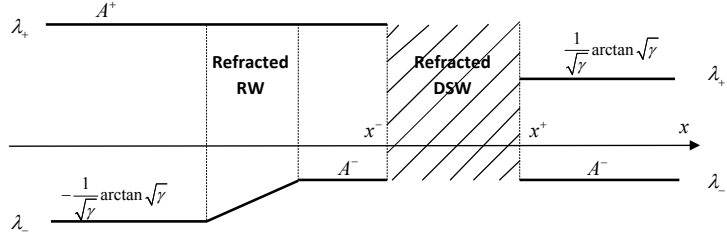}
\caption{Distribution of the dispersionless limit Riemann invariants after the DSW-RW interaction.}
\end{center}\label{fig15}
\end{figure}

Thus, using relationships (\ref{srim}) between the Riemann invariants $\la_{\pm}$ and the hydrodynamic variables $n,u$, one arrives at the set of equations determining  the hydrodynamic states $(n_1, u_1)$ and $(n_2, u_2)$ at the trailing and leading DSW edges respectively:
\begin{equation}\label{125}
\begin{split}
\frac{u_1}{2} + \frac{1}{\sqrt{\gamma}}\arctan \sqrt{\gamma n_1}=A^+; \quad \frac{u_1}{2} - \frac{1}{\sqrt{\gamma}}\arctan \sqrt{\gamma n_1}= \frac{u_2}{2} - \frac{1}{\sqrt{\gamma}}\arctan \sqrt{\gamma n_2}=A^-; \\
\frac{u_2}{2} + \frac{1}{\sqrt{\gamma}}\arctan \sqrt{\gamma n_2}=\frac{1}{\sqrt{\gamma}}\arctan \sqrt{\gamma}\, .
\end{split}
\end{equation}
So
\begin{equation}\label{126}
\begin{split}
n_1=\frac{1}{\gamma} \tan^2\left( \sqrt{\gamma}\frac{A^+ - A^-}{2}\right),  \quad u_1=A^++A^- , \\
n_2=\frac{1}{\gamma}\tan^2\left( \frac{1}{2}\arctan \sqrt{\gamma} - \frac{A^-}{2}\sqrt{\gamma}\right), \quad u_2= A^-+\frac{1}{\sqrt{\gamma}} \arctan{\sqrt{\gamma}} \, .
\end{split}
\end{equation}

To verify our key assumption about the ``semi-classically clean'' DSW-RW interaction in the sNLS equation case we have compared the values of the density  and velocity in the  region between the refracted DSW and RW obtained from direct numerical simulations of the sNLS equation with the predictions for $n_1$ and $u_1$ of formulae (\ref{126}) based on this assumption. As one can see from
Fig.~16 the the comparisons  show an excellent agreement  confirming our clean interaction hypothesis for a range of values of $\gamma$, $A^+$ and $A^-$. At the same, one can notice a small discrepancy visible at larger values of $A^+$ ($A^+ \gtrsim 1.7$) in the plots for $\nu(A^+)$. This is connected with the occurrence of the vacuum point in the refracted DSW for sufficiently large density jumps across it. As was observed in \cite{egkk07}, for large-amplitude photorefractive DSWs the Riemann invariant transition condition (\ref{c1}) is replaced by the classical Rankine-Hugoniot shock jump conditions so relation (\ref{126})  holds for large $A^+$ only approximately.

\begin{figure}[h]
\begin{center}
\includegraphics[width=6.5cm]{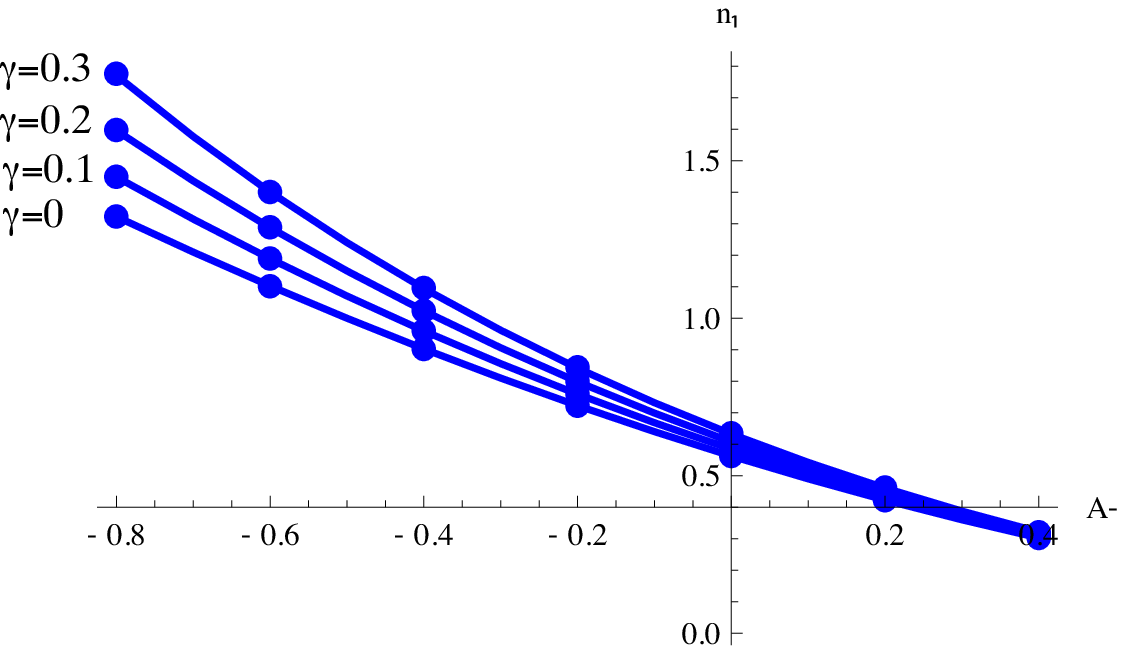} \qquad \quad \includegraphics[width=6cm]{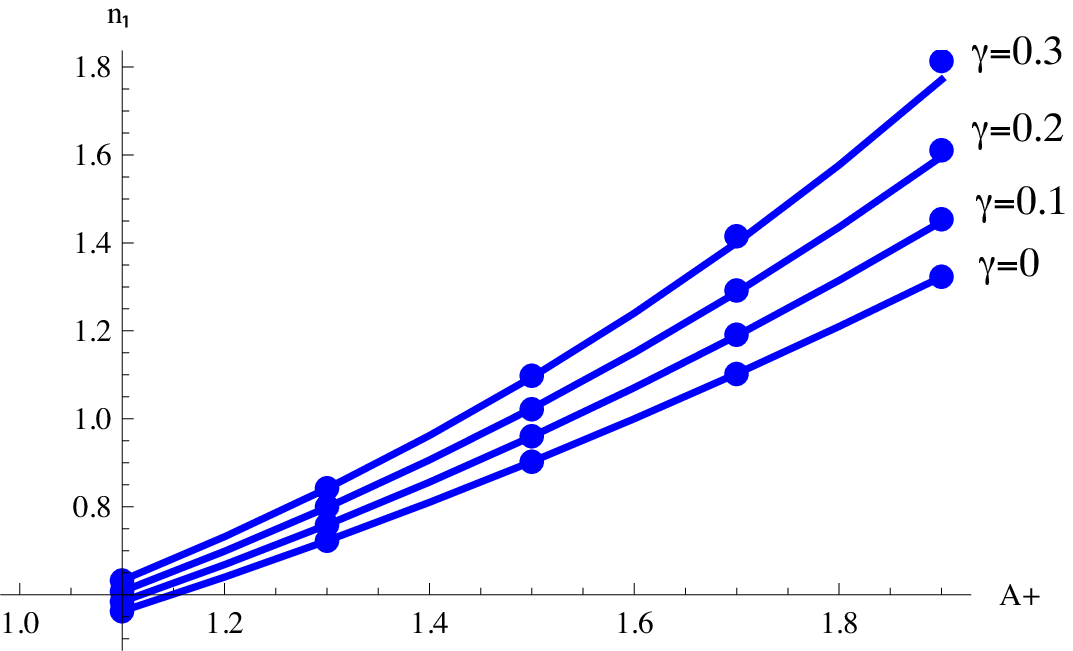}
\caption{Density $n_1$ in the constant flow region between the refracted DSW and RW. Left: $n_1(A^-)$ for fixed $A^+=1.5$; Right: $n_1(A^+)$ for fixed $A^-=0$. Solid lines:
analytic (modulation theory) curves;  dots: direct numerical simulations data.}
\end{center}
\end{figure}
Now, we shall use general relationships (\ref{spm}) --- (\ref{speedamp}) to derive the trailing soliton parameters in the refracted DSW.

Comparing (\ref{c1}) and (\ref{125}) we find  $\la_-^0=A^-$  so expression (\ref{un}) for $\bar u(\bar n)$ assumes the form
\begin{equation}\label{un2}
\bar u(\bar n)= 2 \left(A^- +\frac{1}{\sqrt{\gamma}}\arctan \sqrt{\gamma \bar n}\right)\, .
\end{equation}
Substituting (\ref{un2}) into (\ref{vplus}) and (\ref{Omega}) and using the same  change of variable (\ref{alpha})
in  ODE (\ref{ODEkap}) we arrive at the same ODE (\ref{ODEalpha}) for the function $\widetilde{\alpha}(\bar n)$ but now with a general
boundary condition $\widetilde{\alpha}(n_2)=1$ since $n_2 \ne 1$ for the refracted wave (see (\ref{126})). As before, this condition follows from the boundary condition for $\kappa$ in (\ref{ODEkap}) and the relationship
(\ref{alpha}) between $\widetilde{\alpha}$ and $\kappa$. The velocity
of the trailing soliton in the refracted DSW is determined by Eqs.~(\ref{spm}),
(\ref{Omega}) as
\begin{equation}\label{sminafter}
s^-_r= 2 \left( A^- +\frac{1}{\sqrt{\gamma}}\arctan \sqrt{\gamma n_1} \right)
    +\frac{\sqrt{n_1}}{1+\gamma n_1}\widetilde{\alpha}(n_1)\, ,
\end{equation}
where $n_1$ is now given by Eq.~(\ref{126}).
Comparison for the dependence $s^-_r(A^+)$ for a fixed value of $A^-=-0.8$ is presented in Fig.~18. One can see that the value of $s^-_r$ quite strongly depends on the saturation parameter $\gamma$.
\begin{figure}[htp]
\begin{center}
\includegraphics[width=6cm]{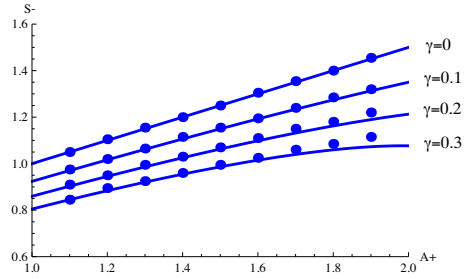}
\caption{The refracted DSW trailing edge speed  $s^-_r$ as a function of an input parameter $A^+$ for fixed $A^-=-0.8$.
Solid lines: modulation solution (\ref{sminafter}); dots: numerical simulations data.}
\end{center}
\end{figure}
Expanding $s^-_r$ for small $\gamma$ we get (cf. (\ref{smin3}))
%\begin{equation}\label{smin4}
%s^-=1+ \frac{A^-+A^+}{2} + \gamma \left[2 \delta^3\left(3\left(\tfrac{\Delta}{\delta} \right)^2+16\tfrac{\Delta}{\delta} - 19 + 64 %\ln \frac{4-\tfrac{\Delta}{\delta}}{3}\right) + \Delta^2(\Delta - 2\delta)+\frac{1}{3}(2\delta^3 -\Delta^3 -1) \right]
%+ O(\gamma^2)
%\end{equation}
\begin{equation}\label{smin4}
s^-_r=1+ \frac{A^-+A^+}{2} + \gamma \left[\frac{2}{3}\Delta^3 + 4\Delta^2 \delta +32 \Delta \delta^2 - \frac{112}{3}\delta^3
+ 128 \delta^3 \ln \frac{4-{\Delta}/{\delta}}{3} - \frac{1}{3}
 \right] + O(\gamma^2)
\end{equation}
Here $\Delta=\frac{A^+ - A^-}{2}$, $\delta = \frac{1-A^-}{2}$.
Again,  one can see that the leading order of expansion (\ref{smin4}) agrees  with the cubic NLS result (\ref{edges}) as expected.

Given the value of $s_r^-$, the trailing dark soliton amplitude $a_s$ in the refracted DSW is  found from formula (\ref{speedamp}). Comparisons of the analytically found values of $a_s$ for $\gamma=0.2$ with direct sNLS numerical simulation data are presented in Fig.~17. and show excellent agreement. Also, the dashed lines  show the dependencies $a_s(A^-)$ and $a_s(A^+)$ for $\gamma=0$. As one can see, the nonlinearity saturation has strong effect on the refracted DSW soliton amplitude.
\begin{figure}[htp]
\begin{center}
\includegraphics[width=5cm]{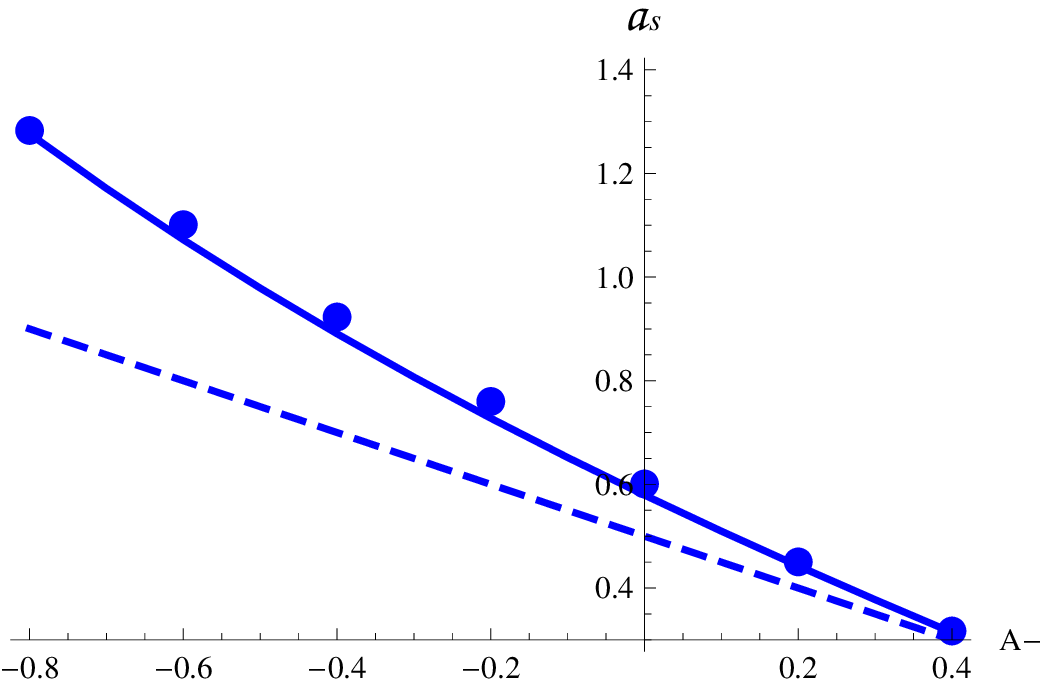} \qquad \quad \includegraphics[width=5cm]{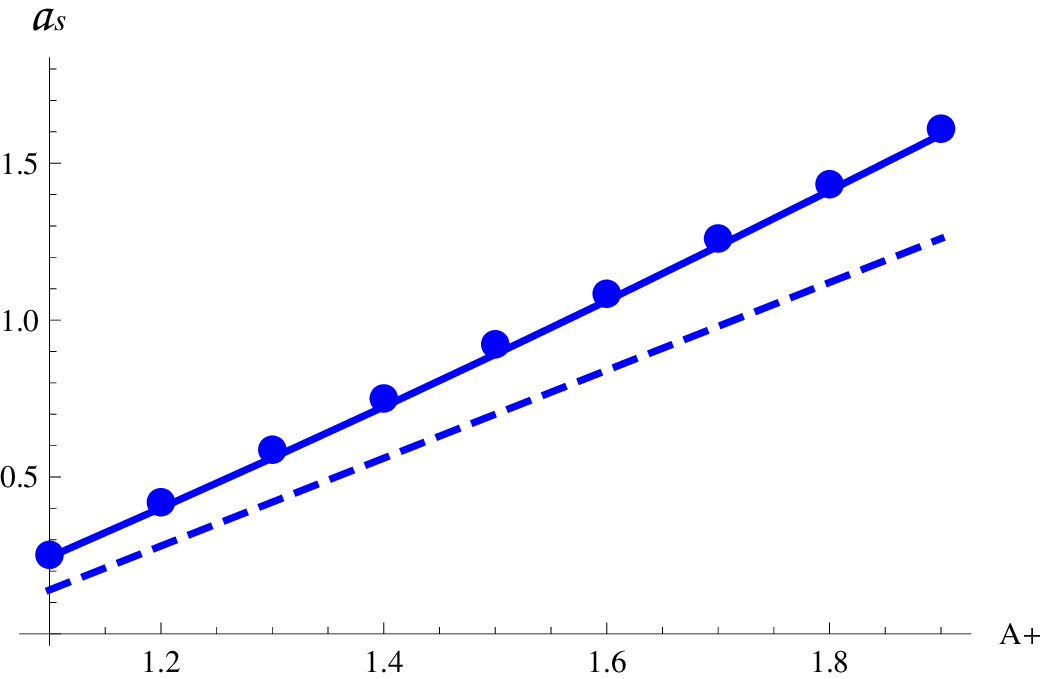}
\caption{Trailing soliton amplitude $a_s$. Left: $a_s(A^-)$ for $A^+=1.5$; Right: $a_s(A^+)$ for $A^-=-0.4$.
Solid line:
analytic curve for $\gamma=0.2$; Dots: direct numerical simulations data for $\gamma=0.2$. Dashed line: the curve for $\gamma=0$.}
\end{center}
\end{figure}
The condition $a_s=n_1$ defining the vacuum point occurrence at the trailing edge of the refracted DSW, leads to the same equation (\ref{vc}), which was obtained earlier for the incident DSW,  with the only (essential) difference that $n_1$ is now given by (\ref{126}). The vacuum point regions diagram for $\gamma=0.2$ is presented in Fig.~19.
\begin{figure}[htp]
\begin{center}
\includegraphics[width=6cm]{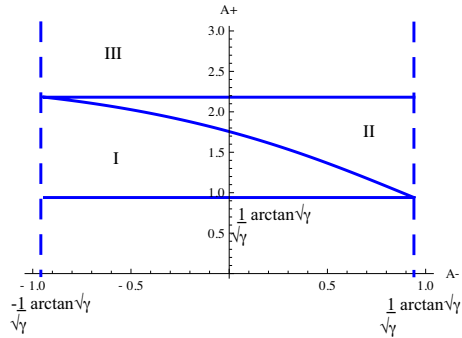}
\caption{Regions of the plane of initial parameters $A^-, A^+$ for $\gamma=0.2$: (I) No vacuum points; (II) No vacuum points in the incident DSW, a vacuum point in the refracted DSW;
(III) Vacuum points in both incident and refracted DSWs.
}
\end{center}\label{fig}
\end{figure}

Comparison with the analogous diagram  for the Kerr nonlinearity case $\gamma=0$ (Fig.~11) shows that  variations of the saturation parameter $\gamma$ have rather significant effect on the
vacuum point appearance. Our numerical simulations confirm this conclusion. As already was mentioned, in the developed modulation theory we assume a semiclassically ``clean'' DSW-RW interaction, which, strictly speaking, applies only to the region I in Fig.~19. However, our comparisons show that, if the initial parameter $A^+$ is not too large, the DSW fitting approach \cite{el05} implementing the Riemann invariant transition condition
(\ref{c1}) gives reasonably good quantitative predictions for the refracted DSW parameters for the regions II and III as well.

\subsection{DSW refraction parameters}
The DSW  amplification coefficient is defined as $\nu=I_r/I_0$, where the incident DSW relative intensity $I_0$ is given  by (\ref{io}).
Using (\ref{126}) the relative intensity of the refracted DSW is readily found  in terms of the input parameters $A^+$ and $A^-$ as (see (\ref{int}))
\begin{equation}\label{ir1}
I_r= \frac{n_1}{n_2}= \frac{\tan^2\left( \sqrt{\gamma}\frac{A^+ - A^-}{2}\right)}{\tan^2\left( \frac{1}{2}\arctan \sqrt{\gamma}-\frac{A^-}{2}\sqrt{\gamma}\right)}\, .
\end{equation}
In Fig.~20
we present the dependencies $\nu(A^-)$ (for a fixed $A^+$) and $\nu(A^+)$ (for a fixed $A^-$). One can see that the amplification coefficient (unlike the individual parameters of the incident and refracted DSWs --- see e.g. Fig.~16 above and Figs.~18, 19 below) shows a very weak dependence on the saturation parameter $\gamma$ for rather broad intervals of $A^+$ and $A^-$ so that one can safely use simple expression (\ref{nu})
obtained for $\gamma=0$. The direct numerical simulations fully confirm this conclusion (we do not present numerical points on Fig.~20 to avoid cluttering the plot).
\begin{figure}[htp]
\begin{center}
\includegraphics[width=5cm]{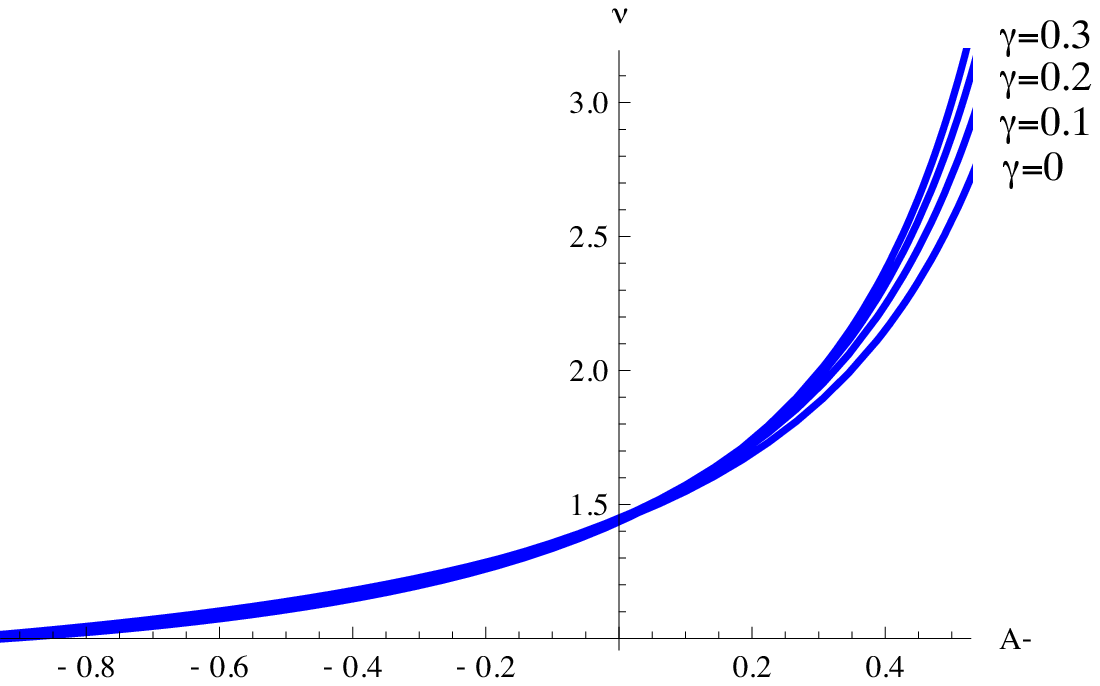} \qquad \quad  \includegraphics[width=5cm]{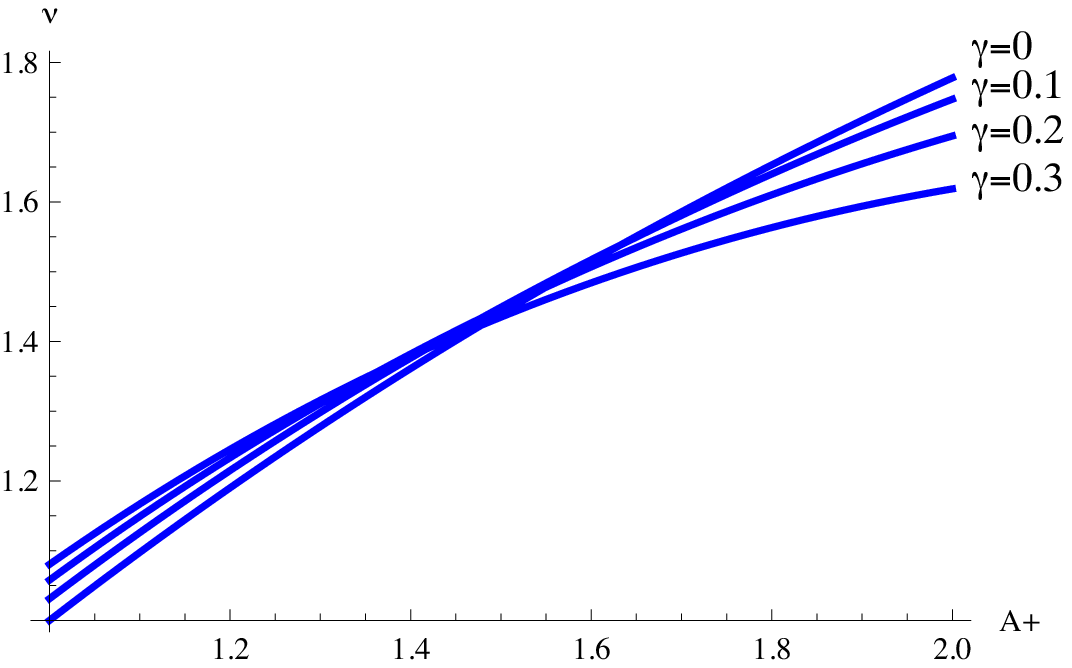}
\caption{DSW amplification coefficient $\nu$. Left: $\nu(A^-)$ at $A^+=1.5$, $A^+=1.5$.  Right: $\nu(A^+)$ at $A^-= 0 $; }
\end{center}
\end{figure}

Now we look at the behaviour of the acceleration coefficient $\sigma=  s^-_r - s^-_0$, which is found analytically with the aid
of formulae (\ref{sminafter}) and (\ref{smin1}). The dependence $\sigma(\gamma)$ for
$A^+=1.2$, $A^-=-0.7$ (Region I in Fig.~19) is shown in Fig.~21.
\begin{figure}[htp]
\begin{center}
\includegraphics[width=5cm]{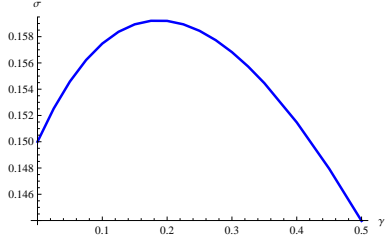}
\caption{Analytical curve for the DSW acceleration coefficient $\sigma$ as a function of the saturation parameter $\gamma$ for
$A^+=1.2$, $A^-=-0.7$. }
\end{center}
\end{figure}
One can see that, similar to the amplification coefficient $\nu$, the dependence of $\sigma$ on  $\gamma$ and $A^+$ (i.e.
on the intensity of the incident DSW) is quite weak.  Indeed, the relative  change of $\sigma$
does not exceed $10\%$ over the broad interval of $\gamma$ from $0$ to $0.5$). Thus, at least in region I, one can safely assume the simple expression
(\ref{sigma}) $\sigma=(1+A^-)/2$ obtained for the cubic nonlinearity case. The comparisons with numerics presented in Fig.~22 confirm this observation.
\begin{figure}[htp]
\begin{center}
 \includegraphics[width=5cm]{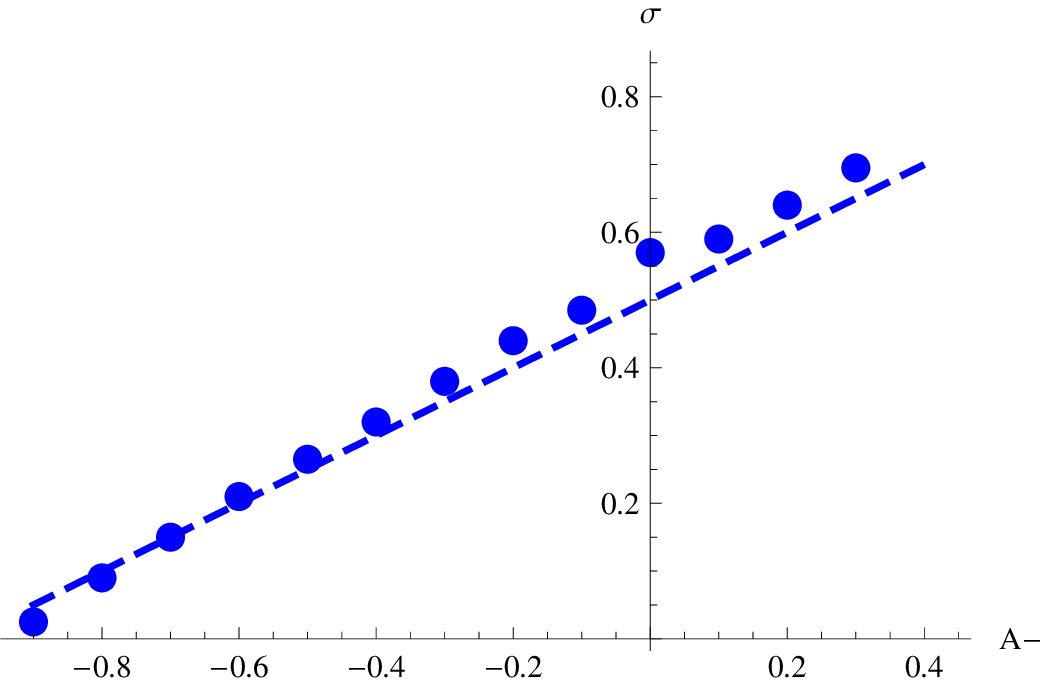} \qquad \includegraphics[width=5cm]{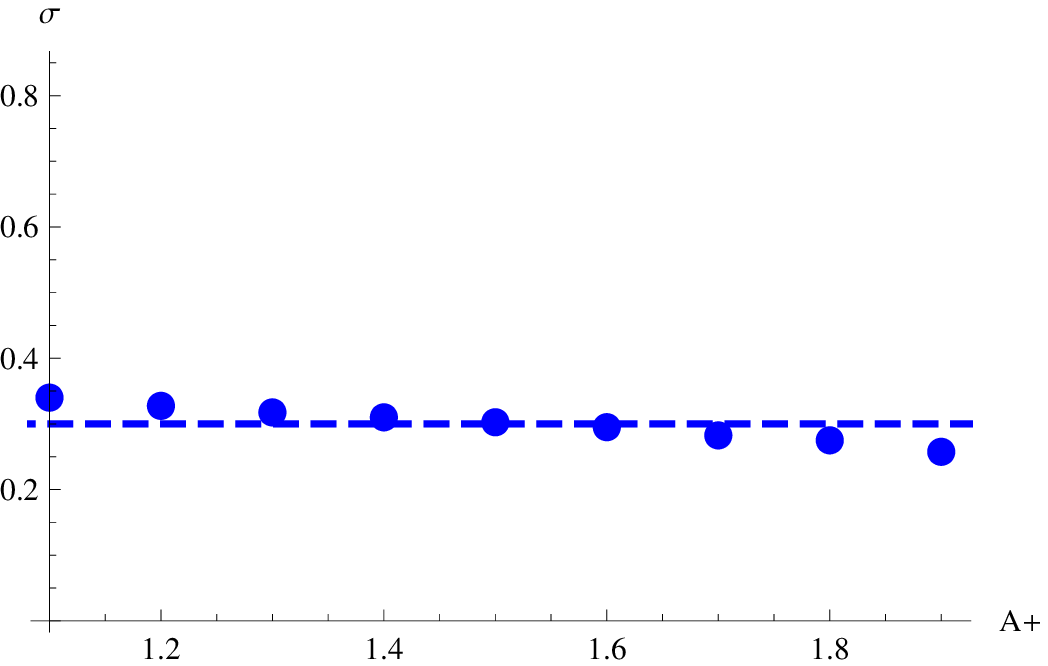}
\caption{The DSW acceleration coefficient $\sigma$ as a function of input parameters $A^-$ and $A^+$. Dashed lines: analytic curves for $\gamma=0$; Cirles:  numerical data for $\gamma=0.2$. Left: $\sigma(A^-)$ at fixed $A^+=1.2$; Right: $\sigma(A^+)$
for fixed $A^-=-0.4$}
\end{center}
\end{figure}
To analytically quantify the deviations of the quite complicated general ``photorefractive'' dependence $\sigma(A^+, A^-, \gamma)$ from the simple dependence $\sigma=(1+A^-)/2$ in the cubic nonlinearity case given by (\ref{sigma}), we derive an asymptotic expansion for $\sigma$ for the case when both interacting waves have small intensity.
Introducing $\varepsilon_+$ and $\varepsilon_-$ by
\begin{equation}\label{}
A^{-}=-\frac{1}{\sqrt{\gamma}} \arctan \sqrt{\gamma} + \varepsilon_-\, , \qquad
A^{+}=\frac{1}{\sqrt{\gamma}} \arctan \sqrt{\gamma} + \varepsilon_+\,
\end{equation}
and assuming $\varepsilon_- \ll 1$, $\varepsilon_+ \ll 1$ we obtain from (\ref{smin3}) and (\ref{smin4}) on retaining second order
terms,
\begin{equation}\label{sigap}
\sigma=s^-_r - s^-_0 = \frac{\varepsilon_-}{2} + \varepsilon_- \gamma + O(\varepsilon_- \gamma^2; \varepsilon_-^2 \gamma; \varepsilon_-\varepsilon_+ \gamma).
\end{equation}
One can see that expansion (\ref{sigap}) does not contain terms proportional to $\varepsilon_+ \gamma$,  which implies that,
for the interactions involving weak photorefractive DSW and RW, the acceleration $\sigma$ of the DSW up to second order does not depend on its initial intensity.

\begin{figure}[htp]
\begin{center}
 \includegraphics[width=5cm]{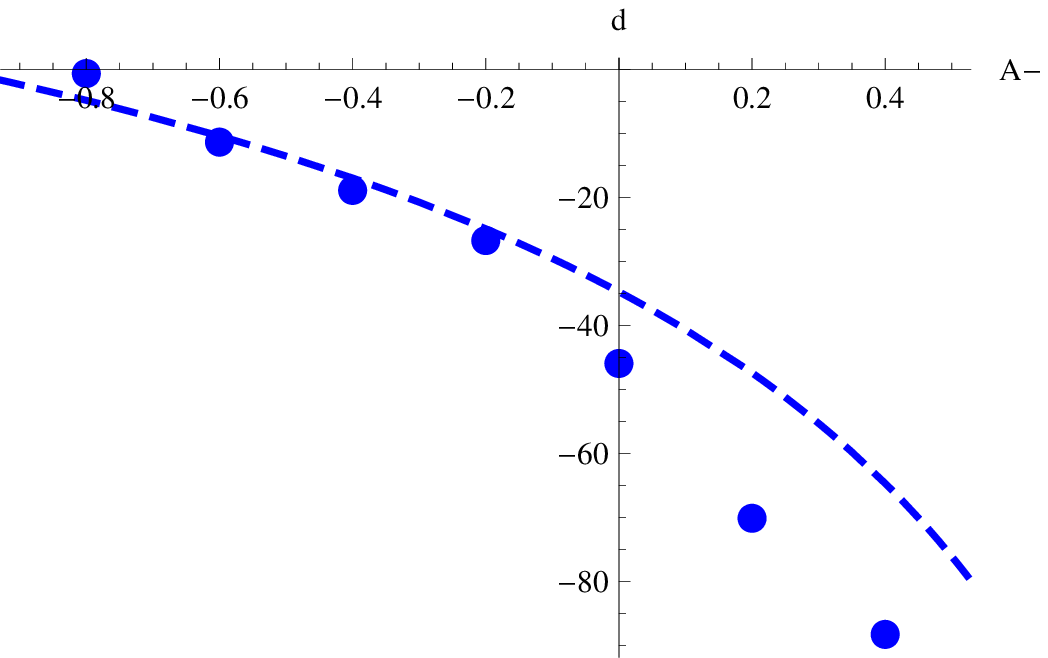} \qquad \includegraphics[width=5cm]{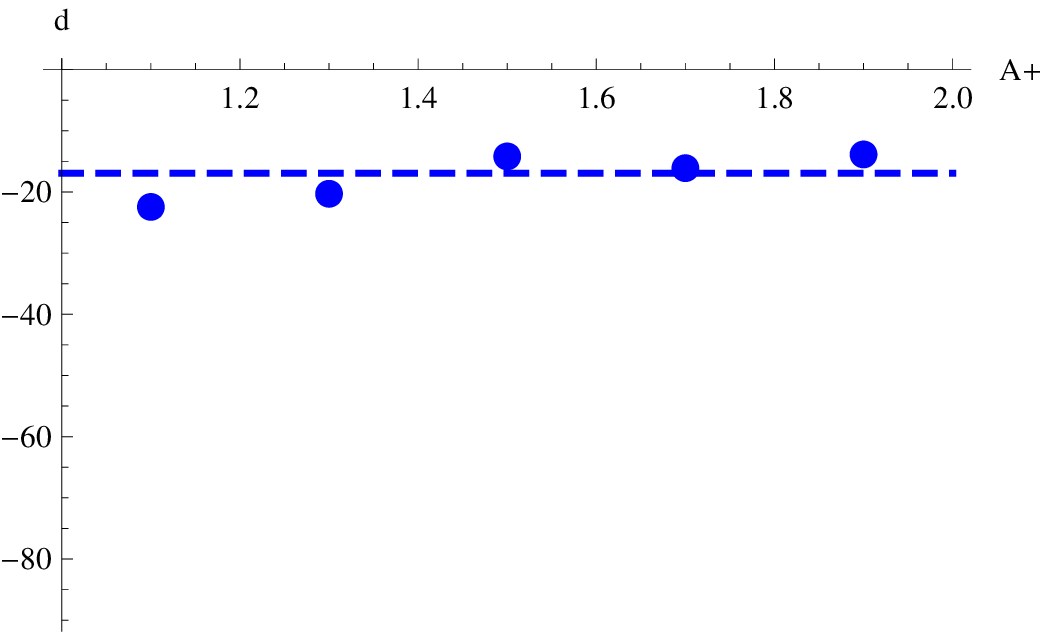}
\caption{DSW refraction phase shift $d$. Left: dependence $d$ on $A^-$ for fixed $A^+=1.5$; Right: dependence $d$ on $A^+$ for fixed $A^-=0$. Dashed lines correspond to $\gamma=0$ (analytical), circles --- to  $\gamma=0.3$ (numerical). }
\end{center}
\end{figure}

Finally, in Fig.~23 we present numerical values for the DSW refraction shift $d$ (see Fig.~4) taken for the particular value of $\gamma=0.3$. The numerics (circles) are put against the analytical curves $d(A^-, A^+)$ defined by formula (\ref{refshift}) for the cubic nonlinearity case, $\gamma=0$. One can see that, similar to other definitive DSW refraction parameters $\nu$ and $\sigma$, there is almost no dependence on $A^+$ and $\gamma$ at a fixed value of $A^-$ (roughly, the RW intensity), however, the departure of the dependence $d$ on $A^-$ from the Kerr case $\gamma=0$ becomes more pronounced with growth of  $A^-$.

\section{Conclusions}

In this paper, we have considered a dispersive counterpart of the classical gas dynamics problem of the interaction  of a
shock wave with a counter-propagating simple rarefaction wave often referred to as the shock wave refraction problem.
Apart from the obvious contrast between both local and global structures of viscous SWs and DSWs,
there is a fundamental difference between the classical dissipative, and the present, dispersive conservative settings which makes possible full quantitative description of the DSW refraction. The salient feature of the viscous SW refraction is the generation of the varying entropy wave resulting in a complicated system of the Rankine-Hugoniot shock conditions  resolvable in most cases only by numerical means. Contrastingly, in conservative dispersive hydrodynamics, the thermodynamic entropy does not change and the jumps of the hydrodynamic quantities across the DSW are completely determined by the transfer of the Riemann invariants of the appropriate modulation Whitham equations along the characteristics. Essentially, the DSW-RW interaction problem reduces, in the semi-classical limit, to the description of the interaction of two expansion fans: one of the shallow-water equations and another one -- of the Whitham modulation equations.

Our study was performed in the frameworks of the one-dimensional defocusing NLS equations with cubic nonlinearity (Eq. \ref{3-1}) and saturable nonlinearity (Eq. \ref{snls}). To model a generic DSW-RW bidirectional interaction we have considered the initial-value problems for both NLS equations with the initial data given by appropriate piecewise-constant distributions for the
density (the wavefunction squared modulus) and the velocity (the wavefunction phase gradient).  To single out the ``pure'' DSW-RW interaction, we  specified the initial data in the form of two steps for the ``Eulerian'' (dispersionless limit) Riemann invariants having jumps of different polarity shifted with respect to one another by a large distance $l$ (see Fig 2a).

For the integrable cubic nonlinearity case we have constructed exact  modulation solutions, asymptotically ($t \gg 1$) describing all stages of the bidirectional DSW-RW interaction in terms of the evolution of the Riemann invariants of the NLS-Whitham system. This was done by mapping the original nonlinear Gurevich-Pitaevskii type matching modulation problem to the Goursat problem for the classical linear Euler-Poisson-Darboux equation (\ref{EPD1}). Along with the modulation solution describing slow variations of the amplitude, the wavelength, the mean etc. in the DSW, we have derived explicit compact expressions for the DSW-RW refraction phase shifts, having certain analogy with the classical soliton phase-shifts in two-soliton collisions.

For the NLS equation with saturable nonlinearity, which is a typical model for the description of the light beam propagation through photorefractive optical materials, we have taken advantage of the DSW fitting method \cite{el05} applicable to non-integrable dispersive systems. This method was applied recently in \cite{egkk07} to the description of the simple-wave optical photorefractive DSWs and in the present study we extended it to the DSW-RW interaction. Our consideration of ``non-integrable'' DSW refraction in the framework of the NLS equation with saturable nonlinearity (\ref{snls}) is based on the assumption (confirmed by direct numerical simulations) that the head-on DSW-RW interaction is ``semiclassically elastic'', i.e. is not accompanied by the generation of new DSWs or/and RWs.
The  comparisons of the key parameters of the photorefractive DSW refraction: the  amplification coefficient $\nu$  and the acceleration coefficient $\sigma$ defined by formulae (\ref{sigma1} a)  and (\ref{sigma1} b) respectively, with their Kerr ($\gamma=0$) counterparts have revealed a rather weak dependence of these particular parameters on the
    saturation coefficient $\gamma$, which could prove useful for the experimental all-optical modelling of the BEC DSW refraction using photorefractive materials.

 The direct numerical simulations of the photorefractive DSW refraction confirm key predictions of our modulation analysis, which provides
 further  striking evidence of the robustness of the modulation theory in non-integrable dispersive wave problems, now in the more complicated setting involving DSW-RW interactions.

We conclude with the remark that the approach used in this paper can also be applied to obtain asymptotic solution to the problem of the {\it overtaking} DSW-RW interaction in the NLS flows. While this problem was studied in the KdV equation framework in \cite{abh09}, we believe that it deserves special attention in the context of the defocusing NLS equation  since, due to a different dispersion sign and the possibility of the vacuum point occurrence within the DSW one can expect a number of qualitative and quantitative differences compared to the KdV flows.  Also, the developed theory can be readily extended to the problem of the generation of DSWs by the nonlinear dispersive interference of two simple rarefaction waves studied numerically in \cite{hec09}.

\subsection*{Acknowledgments}
We are grateful to A. Kamchatnov and M. Hoefer for valuable comments.
V.V.Kh. thanks the London Mathematical Society for partial support of his visit to Loughborough University during which this work was started.

\end{document}